\documentclass[preprint,12pt]{elsarticle}

\usepackage{amssymb}
\usepackage{float}
\usepackage{placeins}
\usepackage[utf8]{inputenc}
\usepackage[margin=2cm]{geometry}
\usepackage{hyperref}

\usepackage{tikz}
\usetikzlibrary{arrows, shapes,decorations.markings,positioning}
\usetikzlibrary{calc,fadings,decorations.pathreplacing,decorations.pathmorphing}
\tikzset{%
  >=latex,
  inner sep=0pt,%
  outer sep=2pt,%
  mark coordinate/.style={inner sep=0pt,outer sep=0pt,minimum size=3pt,
    fill=black,circle}%
}

\usepackage{pgfplots}

\usepackage{amsmath}
\usepackage{caption}
\usepackage{subcaption}
\newcommand{\lra}{\longrightarrow}
\newcommand{\Tr}{\operatorname{Tr}}
\newcommand{\Cof}{\operatorname{Cof}}
\newcommand{\Om}{\Omega}
\newcommand{\R}{\mathbb{R}}

\newcommand{\eps}{\varepsilon}

\renewcommand{\div}{\operatorname{div}}
\newcommand{\I}{\mathbb{I}}
\newcommand{\pa}{\partial}

\newcommand{\revI}{\textcolor{black}}
\newcommand{\revII}{\textcolor{black}}
\newcommand{\revIII}{\textcolor{black}}

\journal{Elsevier}

\begin{document}

\begin{frontmatter}



\title{Fully Eulerian models for the numerical simulation of\\ capsules with an elastic bulk nucleus}


\author[1]{Florian Desmons}
\author[1]{Thomas Milcent}
\author[2]{Anne-Virginie Salsac}
\author[1]{Mirco Ciallella\corref{cor1}}

\address[1]{\'Ecole Nationale Sup\'erieure d'Arts et M\'etiers, Institut de M\'ecanique et d'Ing\'enierie, 33400 Talence, France}
\address[2]{Biomechanics and Bioengineering Laboratory, Universit\'e de Technologie de Compi\`egne, 60203 Compi\`egne, France}
\cortext[cor1]{Corresponding author (\href{mailto:mirco.ciallella@ensam.eu}{mirco.ciallella@ensam.eu})}

\begin{abstract}
In this paper, we present a computational framework based on fully Eulerian models for fluid-structure interaction for the numerical simulation
of biological capsules. The flexibility of such models, given by the Eulerian treatment of the interface and deformations, allows us to easily deal with the large
deformations experienced by the capsule. 
\revII{The modeling of the membrane is based on a full membrane elasticity Eulerian model 
that is capable of capturing both area and shear variations thanks to the so-called backward characteristics.}
In the validation section several test cases are presented with the goal of comparing our results to others present in the literature. 
In this part, the comparisons are done with different well-known configurations (capsule in shear flow and square-section channel),
and by deepening the effect of the elastic constitutive law and capillary number on the membrane dynamics.
Finally, to show the potential of this framework we introduce a new test case that describes the relaxation of a capsule in an opening channel.
In order to increase the challenges of this test we study the influence of an internal nucleus, modeled as a hyperelastic solid, on the membrane evolution. 
\revI{Several numerical simulations of a 3D relaxation phenomenon are presented to provide characteristic shapes and curves related 
to the capsule deformations, while also modifying size and stiffness parameter of the nucleus.}

\end{abstract}



\end{frontmatter}


\section{Introduction}


Biological capsules have revolutionized the fields of medicine and biotechnology, 
offering remarkable advancements and opportunities for research, development, and practical applications.
For example, in medicine, the use of biological capsules have greatly improved drug delivery systems~\citep{del2014biological}. 
By encapsulating drugs within protective coatings, scientists can enhance their stability, solubility, and targeted release. 
Another field of application in biotechnology concerns the encapsulation for the efficient confinement of enzymes within protective matrices, 
enhancing their reusability, and enabling their application in diverse industrial processes~\citep{li2020fabricating}. 

In the context of studying the interaction between such capsules and a surrounding fluid, 
numerical simulation of fluid-structure interaction (FSI) is a very promising tool to understand their behavior
and deformation.
FSI simulations involve the coupled analysis of fluid flow and structural deformation, allowing researchers to investigate the complex 
interactions between structures and their surrounding fluid environments. When applied to biological capsules, 
FSI simulations provide valuable insights into their mechanical behavior, fluid dynamics, and overall performance.
Nowadays, several methods are used to tackle the coupled problem coming from the interaction between fluid and structure.
In the seminal work~\citep{peskin1972flow}, Peskin introduced an innovative approach where Lagrangian markers are 
used to track the membrane deformation. \revII{The elastic force appears as a source term in the fluid equations and 
is spread on the fluid grid with discretized Dirac delta function.}
\revI{This work was established as a reference for future developments of the so-called immersed boundary 
method~\citep{tian2016deformation,mittal2005immersed,lee2003immersed,peskin2002immersed,griffith2020immersed,sotiropoulos2014immersed,doddi2008lateral}}.
\revII{The immersed boundary method is often coupled with Lattice Boltzmann models for the simulation of biological capsules \cite{kruger2014deformability,kruger2014interplay,guglietta2020effects,de2024fluid}. In this case, a finite element solver for the structure is coupled with the Lattice Boltzmann model, which under certain hypotheses converges to the Navier--Stokes equations in the low Mach limit. }

Other popular approaches to solve such problems are based on arbitrary Lagrangian–Eulerian (ALE) 
methods~\citep{liu1998numerical,sahin2009arbitrary,donea1982arbitrary,fernandez2007projection,fernandez2005newton},
which consider body-fitted grids to follow the displacement of the interface. 
\revIII{When compared to the classical Peskin's method, the ALE approach allows for a more precise prediction of the fluid-structure
interface which for immersed boundaries is often spread due to the mismatching between the background mesh and the structure geometry.
This precision can be also maintained when large deformations occur by performing either re-meshing or the more efficient swapping strategies~\cite{alauzet2014changing}.}
Although ALE method has been widely used in many FSI problems, they become extremely cumbersome to apply to 
problems with large deformations, due to the challenging re-meshing process.

More recently, Eulerian models for fluid-structure interaction have been derived for computational purposes as they 
enable complex multi-dimensional applications, with large deformations, to be simulated quite easily on fixed
Cartesian meshes~\citep{cottet2006level,cottet2008eulerian,cottet2021methodes}.
\revIII{The simplicity of this approach lies in the possibility of simulating FSI problems by using classical finite difference 
or finite volume methods on fixed (possibly Cartesian) grids, 
meaning that ideally any fluid solver can be extented to simulate FSI applications. }
\revIII{In contrast with other approaches, however, the fully Eulerian formulation does not allow for the use of 
pre-existing solid models and software, which need to be adapted to the Eulerian framework.}
\revIII{To do so, both the fluid and the solid are described by an Eulerian approach, and modeled as an unique complex flow}.
\revII{Therefore, there is no an actual coupling between a fluid solver and a solid one, contrary to the Lattice Boltzmann--immersed boundary method.}
\revIII{While, interface and deformations of the media are captured by Eulerian fields, and advected by the fluid velocity 
in a level-set fashion~\citep{osher2004level}.} 
In the last decade, this approach has been applied to both incompressible and compressible 
materials~\citep{maitre2009applications,bergmann2022eulerian,gorsse2014simple,de2017cartesian} proving its high flexibility,
especially when dealing with complex geometries.
\revIII{In more recent works~\cite{cottet2016semi,ciallella2023semi,thomann2023implicit}, several authors started working
also on the development of novel numerical approaches to deal with the strict time step constraint imposed by solving fluid and solid 
with a unified solver, which consist in performing an implicit and semi-implicit treatment of the elastic force.}

When dealing with microcapsules, Stokes flows are sometimes considered due to the low Reynolds numbers,  
which corresponds to slow and highly viscous flows. 
Furthermore, the Stokes equations are simpler than the \revII{Navier--Stokes} equations and can be solved more efficiently 
using the boundary integral method~\citep{ramanujan1998deformation,pozrikidis2001interfacial,Lac2004,walter2010coupling}. 
The boundary integral method directly solves the fluid flow problem on the surface 
of the capsule, eliminating the need for a volumetric discretization of the entire fluid domain. 
This reduction in dimensionality significantly reduces the computational cost.
Difficulties remain for the simulation of large deformations, interactions with complex geometries
and applications where the viscosity of the inner fluid is different from that of the outer fluid~\cite{foessel2011influence}.

\revI{The novelty of this paper is to present the application of Eulerian models for FSI for the realistic simulation of biological 
capsules undergoing large deformations with an internal nucleus.} 
In particular, the objective is to study the influence of an internal nucleus, modeled as an elastic solid~\citep{Deborde2020}, 
within the capsule for long-time simulations. The introduction of an internal nucleus is extremely promising for numerous applications
since many biological cells, e.g.\ white blood cells, are often modeled as a thin membrane with a nucleus, also modeled as a thin elastic membrane~\cite{alizad2017numerical}. 
Once the proposed model and method are validated against the literature of microcapsules, 
a new configuration of a relaxation phenomenon~\citep{gires2016transient} is simulated.  
Due to the difficulties in simulating both experimentally and numerically this phenomenon, up to our knowledge this is the first simulation of this kind, 
where the nucleus is modeled as an elastic bulk, and 
it was possible thanks to the flexibility of the fully Eulerian framework.

The paper is organized as follows. 
\revII{First, in Section~\ref{sec:statement} we describe the studied problem, represented by a capsule immersed within an incompressible fluid, and the associated physical parameters.}
Next, in Section~\ref{sec:FSI model}, we briefly recall the formulation of the Eulerian model used to simulated this complex fluid-structure problem.
In Section~\ref{sec:numerics}, we present the general numerical scheme used to discretize the FSI model along with a novel method to mitigate
the degradation of the deformation vector in the case of large deformation and long-time simulations.
The mathematical model, along with the proposed numerical method, is then validated with respect to other approaches present in the literature, in Section~\ref{sec:validation}.
In Section~\ref{sec:simulationbulk}, the relaxation phenomenon is deeply analyzed considering the impact of an internal solid, and its stiffness, on the capsule elastic behavior. 
Finally, in~\ref{sec:circularshear}, we present a test case that is a modification of the sheared elastic sphere relaxation proposed in \citep{Milcent2016}. 

\section{Problem statement}\label{sec:statement}
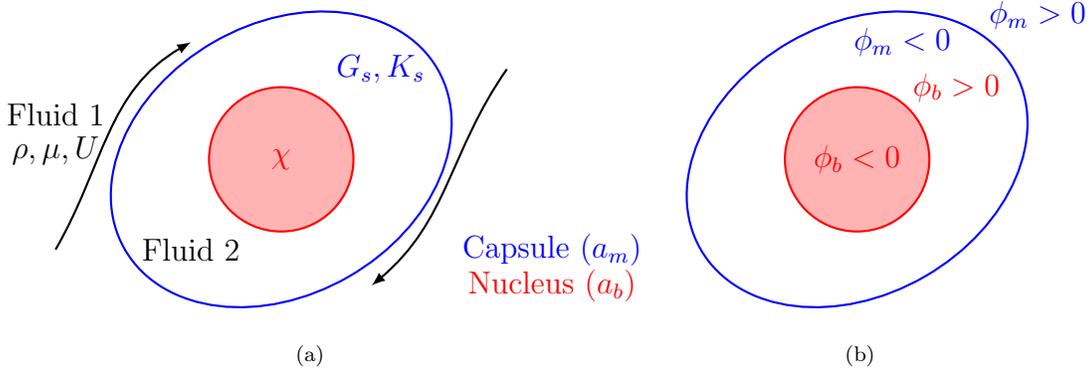
\begin{figure}
\centering
\subfloat[]{
\tikzset{>=latex}
\begin{tikzpicture}[scale=1.2]
         \pgfmathsetseed{1}
         \draw[->,thick] (-2.5,-1) to[in=210,out=60] (-1,1.3);
         \draw[->,thick] (2.5,1) to[in=40,out=235] (1,-1.4);
         \node at (-2.5,0.5) {Fluid 1};
         \node at (-2.5,0.1) {$\rho, \mu, U$};
         \node at (-1,-1) {Fluid 2};
         \draw[rotate=30,thick,blue] (0,0) ellipse (2cm and 1.5cm);
	 \draw[thick,red,fill=red!30] (0,0) circle (0.8cm);
	 \node[red] at (0,0) {$\chi$};
	 \node[blue] at (1.1,1) {$G_s,K_s$};
	 \node[red] at (3,-1.4) {Nucleus ($a_b$)};
	 \node[blue] at (3,-1)   {Capsule ($a_m$)};
\end{tikzpicture}}
\subfloat[]{
\tikzset{>=latex}
\begin{tikzpicture}[scale=1.2]
         \pgfmathsetseed{1}
         \draw[rotate=30,thick,blue] (0,0) ellipse (2cm and 1.5cm);
	 \draw[thick,red,fill=red!30] (0,0) circle (0.8cm);
	 \node[red] at (0,0) {$\phi_b<0$};
	 \node[blue] at (0.5, 1.3) {$\phi_m<0$};
	 \node[blue] at (2.0,1.6) {$\phi_m>0$};
	 \node[red] at (1.1,0.8) {$\phi_b>0$};
\end{tikzpicture}}
\caption{Characteristic variables of the problem.}
\label{fig:variables}
\end{figure}

\revI{We consider an elastic capsule (see Figure~\ref{fig:variables}) modeled as a thin \revIII{massless viscoelastic} membrane, 
with an internal nucleus, modeled as a solid bulk of radius $a_b$ and stiffness $\chi$, immersed within an incompressible
Newtonian fluid with viscosity $\mu$ and density $\rho$. The same viscosity and density are considered in the whole domain.} 
\revIII{The viscosity function is herein treated with a constant value everywhere, meaning that viscous forces also act on the part of the domain occupied by the membrane. By defining the viscosity with appropriate Heaviside functions following \cite{cottet2008eulerian}, it is also possible to consider non-viscous membranes. However, recent works actually propose as effective the addition of physical viscosity to the simulation of biological capsules \cite{guglietta2020effects}.}
The radius $a_m$ of the capsule membrane is taken as a characteristic length of the problem. 
The capsule immersed in the fluid undergoes large deformations and its elastic properties are characterized 
by a \revII{nonlinear} constitutive law depending on the shear modulus $G_s$ and the dilatation modulus $K_s$.
The fluid and the elastic capsule are strongly coupled and thus this phenomenon is a fluid-structure interaction problem.
Therefore, the phenomena arising are governed by two dimensionless numbers: the Reynolds number 
\begin{equation}
Re = \frac{\rho U a_m}{\mu},
\end{equation}
and the capillary number
\begin{equation}
Ca = \frac{\mu U}{G_s} 
\end{equation}
where $U$ is the fluid reference velocity.\\ 
The additional dimensionless parameters $a_b/a_m$ and $\chi/G_s$ are considered to study the effect of the
internal nucleus on the capsule profile and deformation. 
In this work, we propose the original idea of using a fully Eulerian model for the simulation of capsules. 
\revII{In this context, the fluid-structure interfaces for membrane and bulk are represented as the zero-level-sets of the Eulerian fields $\phi_m$ and $\phi_b$, defined as signed distance functions (see Figure \ref{fig:variables}b). For more information on the level-set method, we refer to the seminal paper~\cite{Osher1988}.}
The elastic deformations of the capsule are described by Eulerian fields that are advected by the fluid velocity 
and the associated elastic force appears as an Eulerian source term in the fluid equations.

\section{Fully Eulerian fluid-structure model}\label{sec:FSI model}
\subsection{Eulerian description of the fluid}
\revII{Herein, the fluid is modeled as incompressible and Newtonian and the flow evolution is described by the Navier--Stokes 
equations written in their incompressible form:}
\begin{equation}
\left\{
\begin{aligned}
\rho ( \pa_t u + (u\cdot\nabla) u ) -\div(2\mu D(u))+ \nabla p &= F \\
\div(u) &= 0
\label{eq_NS}
\end{aligned}
\right.
\end{equation}
where $D(u) = \frac{1}{2}([\nabla u] +[\nabla u]^T)$ is the strain rate tensor, $p$ the pressure and $F$ is a source term.\\
When no structure is considered, the source term $F$ is set to zero retrieving the original \revII{Navier--Stokes} equations.
In later sections, more details are given on the formulation of the elastic force term for both thin membranes and elastic solids.
\subsection{Eulerian description of solid deformations}
\label{initial}
\begin{figure}
\centering
\tikzset{>=latex}
\begin{tikzpicture}[scale=1.2]
	 \pgfmathsetseed{1}
         \draw (-0.2,2) node[left] {$Y(x,t)=\xi$};
         \fill [black] (0,2) circle (2pt);
         \draw (5.7,2.5) node[right] {$x=X(t,\xi)$};
         \fill [black] (5.5,2.5) circle (2pt);
	 \draw[->,thick] (5.5,2.5) -- (4.8,3.5);
	  \draw (4.8,3.7) node {$n(x,t)$};
         \draw[->,thick,dashed] (0,2) to[bend left] (5.5,2.5);
         \draw[->,thick,dashed] (5.5,2.5) to[bend left] (0,2);
	 \draw (-2,-1.5) node[below right] {Initial configuration $\Omega_0$};
	 \draw (5,-1.5) node[below right] {Deformed configuration $\Omega_t$};
	\draw[very thick] (0,1.4) circle (60pt);
\draw[very thick] (8,0) .. controls +(0.8,1.2)   and +(0.7,-0.7) ..
(7,3.5)
                         .. controls +(-0.9,0.6) and +(0.5,0.6) .. (5,3)
                         .. controls +(-0.6,-0.6)  and +(-0.4,0.6) ..
(5.5,0)
                         .. controls +(0.4,-0.5)  and +(-0.7,-1.3) ..
(8,0) -- cycle;

\end{tikzpicture}
   \caption{Forward and backward characteristics.}
\label{characXY}
\end{figure}
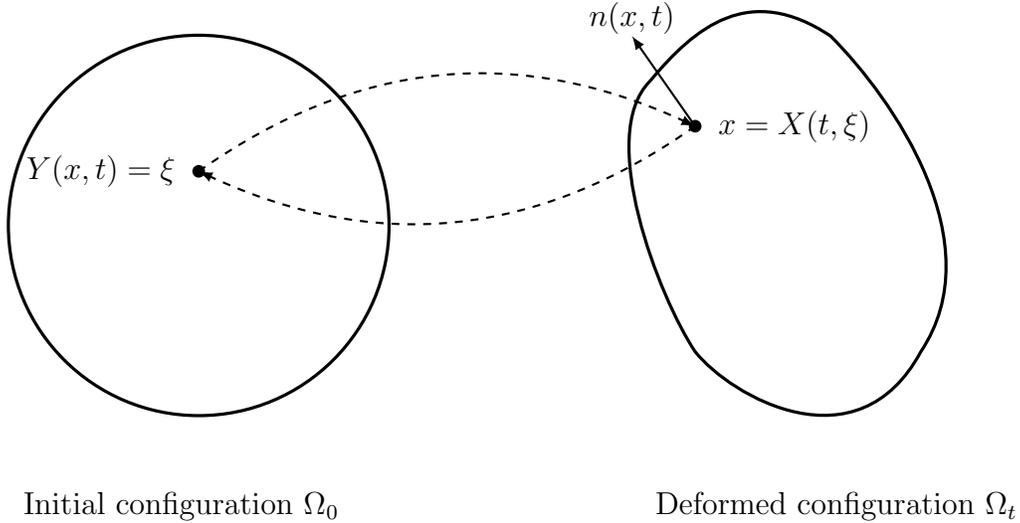

Let $\Om_0\subset \R^3$ be the reference configuration of a
continuous medium and  $\Om_t\subset \R^3$ the deformed configuration at
time $t$. In order to describe the evolution of this medium in the
Lagrangian frame we define the forward characteristics  $X(t,\xi)$ as
the image at time $t$ in the deformed configuration $\Om_t=X(t,\Om_0)$ of a material point
$\xi$ belonging to the initial configuration, i.e., $X:\R^{+}\times \R^{3}\lra \R^3$
 (see Fig.~\ref{characXY}). The corresponding Eulerian velocity field is
defined as   $u:\R^3\times\R^{+}\lra \R^3$ where
\begin{equation}
\partial_tX(t,\xi) = u (X(t,\xi),t),  \qquad   X(0,\xi)=\xi, \qquad \xi\in\Om_0.
\label{caract}
\end{equation}
To describe the continuous medium in the Eulerian frame, we introduce
the backward characteristics  $Y(x,t)$
that for a time $t$ and a point $x$ in the deformed configuration, gives
the corresponding initial point $\xi$ in the initial configuration,
i.e., $Y:\R^3\times \R^{+}\lra \R^3$ (see Fig.~\ref{characXY}).
Since $Y(X(t,\xi),t)=\xi$, differentiating with
respect to time and space  in turn we have:
\begin{equation}
\partial_tY + (u\cdot \nabla_x) Y = 0,  \qquad  Y(x,0)=x, \qquad x\in\Om_t,
\label{transport}
\end{equation}
and
\begin{equation}
[\nabla_{\xi}X(t,\xi)] = [\nabla_x Y(x,t)]^{-1}, \quad\text{for }x=X(t,\xi).
\label{grad_XY}
\end{equation}
The relation (\ref{transport}) is the Eulerian equivalent of the characteristic equation (\ref{caract}). In addition, equation (\ref{grad_XY}) allows to compute the gradient of the deformation in the Eulerian frame via $Y$. 
\revIII{The deformations $Y$ are defined as Eulerian fields in the whole computational domain, 
and are then used to compute the elastic force only in the vicinity of fluid-structure interface.}
The next sections are devoted to the description of bulk and membrane elastic deformations with the backward characteristics $Y$.

%
%
\subsection{Hyperelastic bulk models}
%
%
For hyperelastic bulk material, the internal energy $\mathcal{E}_b$ is a function of the deformation tensor $\nabla_{\xi} X$. We focus in this paper on materials that are Galilean invariant and isotropic. With these assumptions it can be proven (see \citep{holzapfel00}) that $\mathcal{E}_b$ is expressed as a function of the invariants of the left Cauchy-Green tensor $B = [\nabla_{\xi}X][\nabla_{\xi}X]^{T}$. This tensor is written in the Eulerian description with (\ref{grad_XY})
\begin{equation}
B(x,t) = [\nabla_{x}Y]^{-1}[\nabla_{x}Y]^{-T}.
\label{eqB}
\end{equation}
In this work, we consider the neo-Hookean constitutive law ($\mathcal{E}_b$ depends only on $\Tr(B)$). The associated Cauchy stress tensor is given by $\sigma_b=2 \chi B$ where $\chi$ is the elastic modulus coefficient. In this work, we aim at developing a numerical tool able to simulate elastic capsules with a bulk kernel inside. The neo-Hookean model allows to describe, with one parameter, elastic media subjected to large but moderate deformations. To model various aspects of the resulting \revII{nonlinear} relation  between stress and deformation, other elastic constitutive laws can be considered: \revII{Mooney--Rivlin}, \revII{Saint Venant--Kirchhoff}, Ogden. In the final model the bulk solid will be immersed in a incompressible fluid and the fluid-structure interface will be captured by a level set function $\phi_b:\R^3\times \R^{+}\lra \R$ which is advected by the Eulerian velocity field $u$:
\begin{equation}
\partial_t \phi_b + u\cdot \nabla \phi_b = 0.
\label{eq:phibulk}
\end{equation}
The force associated to the neo-Hookean solid is then given by
\begin{equation}
F_b = \div\left(  H\left(\frac{\phi_b}{\varepsilon}\right) \sigma_b \right),  \qquad \sigma_b = 2 \chi B,
\label{eqFb}
\end{equation}
where $H$ is a smoothed Heaviside function and $2\varepsilon$ is the width of the interface. Therefore $\phi_b>\varepsilon$ corresponds to the fluid domain and $\phi_b<-\varepsilon$ to the solid domain.
\subsection{Hyperelastic membrane models}
The notations and results summarized in this section are detailed in \citep{Milcent2016}. We consider a surface $\Gamma_t =\{ x \in \R^3 \; / \; \phi_m(x,t)=0 \}$ captured by a level set function $\phi_m:\R^3\times \R^{+}\lra \R$ and advected by the Eulerian velocity field $u$:
\begin{equation}
\partial_t \phi_m + u\cdot \nabla \phi_m = 0.
\label{AB}
\end{equation}
The normal $n(x,t)$ for  $x\in\Gamma_t$ is then expressed in terms of the normalized gradient of the level set:
\begin{equation}
n(x,t) = \frac{\nabla \phi_m(x,t)}{|\nabla \phi_m(x,t)|}.
\label{normales}
\end{equation}
 To measure the deformations on the surface $\Gamma_t$ we introduce the tensor
\begin{equation}
 \mathcal{A} = B - \frac{(Bn)\otimes(Bn)}{(Bn)\cdot n}.
\label{AB}
\end{equation}
where $B$ is given by \eqref{eqB}. This tensor measures surface deformations by projecting the 3D deformations (measured by $B$) on the surface $\Gamma_t$ (represented locally by $n$). The vector $n$ is an eigenvector of $\mathcal{A}$ associated to the eigenvalue $0$ so $\det(\mathcal{A})=0$. We denote by $(\lambda_1)^2$ and $(\lambda_2)^2$ the two other eigenvalues. The other invariants are used to define the following quantities
\begin{align}
Z_1 &= \sqrt{\Tr(\Cof(\mathcal{A})) } = |\lambda_1 \lambda_2|, \\  Z_2 &= \frac{\Tr(\mathcal{A})}{ 2 \sqrt{\Tr(\Cof(\mathcal{A}))}}= \frac{1}{2} \left( \left|\frac{\lambda_1}{\lambda_2}\right| + \left|\frac{\lambda_2}{\lambda_1}\right| \right) .
\label{Z1_Z2b}
\end{align}
The eigenvalues $\lambda_1$ and $\lambda_2$ correspond to the local deformation of the surface, therefore it is intuitive that $Z_1$ measures the local area variation whereas $Z_2$ measures the local shear variation. These results are demonstrated in details in \citep{Milcent2016}. Note also that $Z_1\geq 0$ whereas $Z_2\geq 1$. Now we introduce a constitutive law $E$ that depends on the invariants $Z_1$ and $Z_2$ and the associated  energy which is localized in a neighborhood of the membrane,
\begin{equation}
\mathcal{E}_m = \int_{Q} E(Z_1,Z_2)\frac{1}{\eps}\zeta\left(\frac{\phi_m}{\eps}\right)\;{\rm d}x.
\end{equation}
Here $Q$ is a box containing the membrane, $\varepsilon$ the width of the interface and $\zeta$ is a cut-off function used to spread the interface near $\{ \phi_m=0\}$. 
\revI{There is no unique way to define such cut-off functions and for this reason, we discuss it in the numerical part where we present the function we decided to use.} 
\revIII{The use of the mollification given by the cut-off function $\zeta$ is necessary to discretize the Dirac delta function contained in  the mathematical model. This is related to the fact that the background fluid mesh does not align with the structure geometry, as also discussed in \cite{peskin2002immersed}. More insights on the mathematical and asymptotic analysis of Eulerian models for FSI is provided in
\cite{cottet2006level,cottet2008eulerian}, where the interface width is discussed to study the convergence of the model.}

Elastic energy is conservative so a variation of the membrane shape induces an elastic force (principle of virtual power) and is given by
\begin{equation}
F_m = \div\left( E_{,1}(Z_1,Z_2) Z_1 \mathcal{C}_1 + E_{,2}(Z_1,Z_2) Z_2 \mathcal{C}_2 \right) \frac{1}{\eps}\zeta\left(\frac{\phi_m}{\eps}\right),
\label{eqFm}
\end{equation}
where
\begin{equation}
\mathcal{C}_1 = \I - n\otimes n, \qquad   \mathcal{C}_2 = \frac{2\mathcal{A}}{\Tr(\mathcal{A})} - (\I - n\otimes n),
\label{Ci}
\end{equation}
and $E_{,i}$ represents the derivative of $E$ with respect to $Z_i$. Three constitutive laws to model membranes are used in this article:
\begin{itemize}
  \item \revII{Evans--Skalak} model:
\begin{equation}
        E_{,1}(r_1,r_2) = K_s (r_1-1), \qquad     E_{,2}(r_1,r_2) = G_s,
    \label{eq:law_evan_skalak}
\end{equation}
\item Skalak model:
\begin{equation}
        E_{,1}(r_1,r_2) = \frac{G_s}{2} (-r_{1}^3+4 r_{1} r_{2}^2 - r_{1} -2r_{2}) + \frac{K_s}{2}(r_{1}^3-r_{1}),\qquad E_{,2}(r_1,r_2) = G_s r_{1} (2r_{1}r_{2}-1),
    \label{eq:law_skalak}
\end{equation}
\item Membrane \revII{Neo--Hookean} model:
\begin{equation}
        E_{,1}(r_1,r_2) = G_s\left(r_2 - \frac{1}{r_{1}^3}\right),  \qquad  E_{,2}(r_1,r_2) = G_s r_1.
    \label{eq:law_neo_hookean}
\end{equation}
\end{itemize}

\subsection{Overall model}
The elastic membrane and bulk media are immersed in a incompressible fluid modeled by the \revII{Navier--Stokes} equations. The overall fully Eulerian model is given by
\begin{equation}
\left\{
\begin{aligned}
\rho ( \pa_t u + (u\cdot\nabla) u ) -\div(2\mu D(u))+ \nabla p &= F_b(\phi_{b},Y_{b}) + F_m(\phi_{m},Y_{m})\\
\div(u) &= 0\\
\pa_t Y_{\ell} + (u\cdot \nabla) Y_{\ell} &= 0   \qquad  \ell = b,m\\
\pa_t \phi_{\ell} + u\cdot \nabla \phi_{\ell} &= 0  \qquad  \ell = b,m
\label{eq_tot}
\end{aligned}
\right.
\end{equation}
The membrane force $F_m$ is given by \eqref{eqFm} and the bulk force $F_b$ is given by \eqref{eqFb}. Note that a level set function $\phi_{\ell}$ and a backward characteristic field $Y_{\ell}$ is needed for each media (bulk ($b$) and membrane ($m$)). 
\revIII{In this work, we always considered $\mu$ as constant in the whole computational domain. However, this model can be easily adapted
to deal with varying viscosity coefficient by taking $\mu$ as a function of $\phi_m$, as also shown in~\cite{cottet2008eulerian}.}
These equations are completed with appropriate initial and boundary conditions that will be detailed in the numerical results sections.
%
%
%
%
%
%
\section{Numerical Scheme}\label{sec:numerics}
%
\subsection{General scheme}
\begin{figure}[!h]
\centering
\begin{tikzpicture}[scale=2.5]
\node[mark size=3pt,color=black] at (0,0) {\pgfuseplotmark{square*}};
\draw (-0.2,-0.1) node[below right] {$p,Y,\phi$};
\draw (-1,-1)--(-1,1);
\draw (1,-1)--(1,1);
\draw (-1,-1)--(1,-1);
\draw (-1,1)--(1,1);
\node[scale=2] at (1,0) {$>$};
\node[scale=2] at (-1,0) {$>$};
\draw (0.8,-0.1) node[above] {$u_1$};
\node[scale=2] at (0,1)  {$\wedge$};
\node[scale=2] at (0,-1) {$\wedge$};
\draw (0.1,0.8) node[left] {$u_2$};
\draw[<->] (1.2,-1)--(1.2,1);
\draw[<->] (-1,1.2)--(1,1.2);
\draw (1.25, 0) node[right] {$\Delta x_2$};
\draw (0,1.25) node[above] {$\Delta x_1$};
\end{tikzpicture}
\caption{Staggered grid with position of unknowns}
\label{mac}
\end{figure}
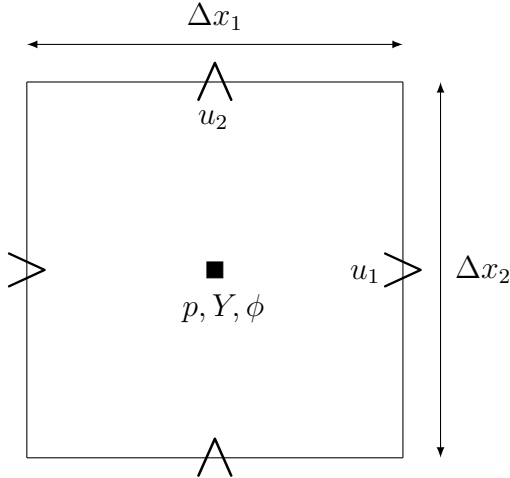

The equations (\ref{eq_tot}) are discretized with finite volume schemes on a staggered grid (see Figure \ref{mac} for a 2D configuration). In the following, the notation $\phi$ (respectively $Y$) denote either $\phi_b$ or $\phi_m$ (respectively $Y_b$ or $Y_m$) because the same discretization is used for each variable.
Let $\Delta t$ be the time step and $u^n,p^n,\phi^n,Y^n,\rho^n,\mu^n$ the time discretization of the variables at $t_n=n\Delta t$. The semi-discretization in time is given by the projection method (see \citep{Guermond2006} for an overview on projection methods) for the \revII{Navier--Stokes} equations (Steps 1-2-3) and an Euler explicit scheme for the advection equations (Step 4):
\begin{align*}
\text{Step 1}:&\rho^n \left(\frac{u^{\star}-u^{n}}{\Delta t} + \div(u^n\otimes u^{\star})\right) -\div(2\mu^n D(u^{\star}) ) + \nabla p^n = F(\phi^n,Y^n) \\
\text{Step 2}:&\div\left(\frac{\Delta t}{\rho^n} \nabla\psi^{n+1}\right) = \div(u^{\star})\\
\text{Step 3}:&u^{n+1} = u^{\star} - \frac{\Delta t}{\rho^n}\nabla \psi^{n+1},  \qquad  p^{n+1} = p^{n} + \psi^{n+1}  \\
\text{Step 4}:&\frac{\phi^{n+1}-\phi^n}{\Delta t} + u^{n+1}\cdot \nabla \phi^n = 0,\qquad \frac{Y^{n+1}-Y^n}{\Delta t} + (u^{n+1}\cdot \nabla) Y^n = 0.
\end{align*}
In Step 1 a prediction of the velocity $u^{\star}$ is computed with an Euler implicit scheme for the viscous term. The convection term is treated explicitly in time with \revII{high-order Runge--Kutta scheme} \citep{ketcheson2008highly} and in space with a \revII{weighted essentially non-oscillatory (WENO5) reconstruction} \citep{shu1996weno}, following an high-order momentum procedure \citep{desmons2021generalized}, and the source term explicitly with a standard second-order discretization. In step 2, the Poisson equation for the pressure increment $\psi^{n+1}$ is solved with appropriate boundary conditions \citep{Guermond2006}. The resulting linear systems are solved with the GMRES algorithm of the HYPRE library \citep{Hypre2002,Hypre2006} with Jacobi preconditioning for Step 1 and with multigrid preconditioning for Step 2. In Step 3 the velocity is corrected to enforce the incompressibility condition and the pressure is updated. In Step 4 the transport equations are discretized again with an explicit high-order Runge-Kutta scheme in time and a WENO5 scheme in space for $\phi,Y$.

For the cut-off function, we considered the following expression $\zeta(r)=\frac{1}{2}(1+{\rm cos}(\pi r))$ on $[-1,1]$ and $\zeta(r)=0$ elsewhere. The Heaviside function is defined as $H(r) = \int_{-\infty}^r \zeta(x) dx$. In our simulations $\varepsilon$ is fixed at $2 \Delta x$ which is the standard value used in the literature to spread the interface.

\subsection{Extrapolations, reinitilization and inner diffusion}\label{sec:aslamdiffusion}

It was already discussed in section 3.2 of \citep{Deborde2020} that the numerical scheme associated  to the fully Eulerian fluid-structure model with bulk (corresponding here to the terms indexed by $b$ in \eqref{eq_tot}) can be unstable in severe situations. Indeed the backward characteristics $Y$ and the level set $\phi$ are computed on the whole domain and these fields can become distorted and irregular in the vicinity of the fluid-structure interface. These perturbations can lead to numerical errors on the location and magnitude of the elastic force and therefore create instabilities. These limitations were overcome using a redistancing technique for the level set function \citep{Sussman1994,Russo2000} together with a linear Aslam extrapolation \citep{Aslam2003} on the backward characteristics of the external fluid.

\begin{figure}
    \centering
    \subfloat[]{\includegraphics[width=80mm]{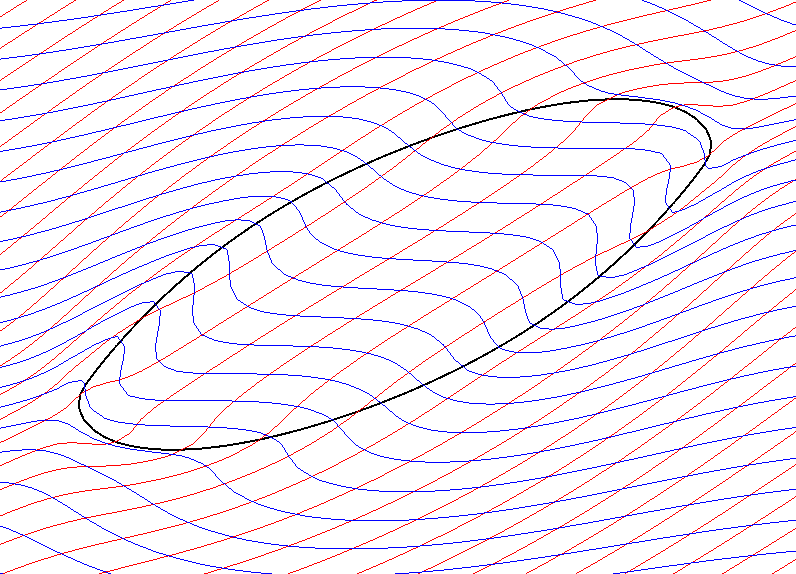}}\qquad
    \subfloat[]{\includegraphics[width=80mm]{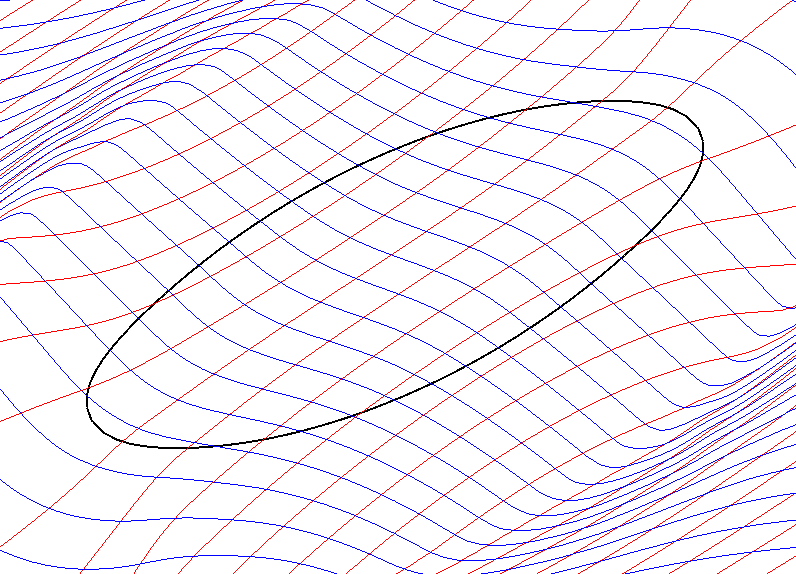}}
    \caption{Deformation of the isolines of $Y$ at early stage rotation in the linear shear illustration: (a) without Aslam extrapolations and inner diffusion; (b) with Aslam extrapolations and inner diffusion.}
    \label{fig:test1}
\end{figure}

In numerical simulations of fluid-structure interaction with bulk materials, the backward characteristics can become distorted only in the outside fluid region. In the inside region, the characteristics are smooth because the elastic forces tend to bring them back to their initial configuration.
However these characteristics can become irregular in the interior fluid in the case of a thin elastic membrane surrounded from either side by a fluid. To illustrate this property we consider a representative test case in capsule applications where an elastic membrane is immersed within a fluid where a linear shear configuration is considered. More information about the numerical simulation of this test case are given in Section \ref{sec:linearshear}. The membrane will elongate in one direction and then will rotate along itself in a stationary velocity field: this behavior is called the tank-treading motion. The isolines of $Y$ are plotted in Figure \ref{fig:test1}a at the early stage of rotation. It is clear that for larger physical times, i.e when the membrane will perform several full rotation, the characteristics will become even more distorted.
This eventually leads also to a bad representation of the membrane.

The first idea to address this problem was to use an Aslam extrapolation of $Y$ inside the fluid with the information coming from the membrane. Since the membrane is closed, the characteristics will cross together and then give rise to a non smooth field. Instead, we propose in this paper to perform a method, we called \textit{inner diffusion}: the backward characteristics are smoothed by using a diffusion equation in the internal region. In Figure \ref{fig:test1}b we present the same illustration where Aslam extrapolations, and the inner diffusion  were used. We clearly see that the characteristics are smooth everywhere and this allows us to run more severe configurations for longer times.

For the reinitilization technique we solve the equation for a fictitious time $\tau$,
$$\pa_\tau \varphi + {\rm sgn}(\varphi_0)(|\nabla \varphi| -1) = 0 $$ 
with a WENO5 scheme.
For the linear Aslam extrapolation, 
$$\pa_\tau Y_n + H(\varphi) n\cdot \nabla Y_n = 0,\quad \pa_\tau Y + H(\varphi) (n\cdot \nabla Y - Y_n) = 0$$
the normal derivative of $Y$ is extrapolated firstly in a constant, and then in a linear,
manner in the region $\varphi>0$ ($H$ is a Heaviside function) 
with the initial condition $Y_n(t=0)=n\cdot \nabla Y$. 
Both equations are discretized by an Euler explicit scheme in time and a WENO5 scheme in space.

For the inner diffusion approach, we solve the equation $\pa_\tau Y = \Delta Y $, in the region $\varphi<0$, by discretizing it with an Euler explicit scheme in time and a centered second-order scheme in space.
The fictitious time $\tau$ has been set for every test case depending on the configurations: more severe tests may need more diffusion. 

\revIII{Similarly to the level-set reinitialization, Aslam extrapolation and inner diffusion avoid to have spurious errors on $Y$
propagating to the fluid--structure interface and altering the computation of the elastic force, along with the problem dynamics.
However, as for the reinitialization, the use of such tools should be restricted only to when it is needed, because the excessive
use may bring unwanted diffusion in the problem. For this reason, the use of extrapolations, reinitilization and inner
diffusion should be adapted to the considered test case. In our simulations, we perform extrapolations, reinitilization and inner diffusion once every ten Navier–Stokes iterations.}

\section{Validation}\label{sec:validation}

In this section, two test cases have been setup with different physical parameters (size, elasticity) of the membrane
in order to validate the proposed method with respect to other approaches present in the literature.
One additional test case, relevant for the grid convergence of the proposed approach has been proposed in~\ref{sec:circularshear} 
starting from a similar test introduced in~\cite{Milcent2016}. 
\revIII{In particular, \ref{sec:circularshear}  serves also the purpose of showing the numerical convergence of the present
approach with a set of three refined mesh, and setting an acceptable mesh refinement to avoid an unreasonable computational time. 
In all numerical simulations present in the core of the manuscript, the  accuracy provided by the second level of refinement 
has been taken as a reference to also provide a good resolution of the flow when the capsules get very close to the wall.}

In particular, this academic test is shown to give an idea about the required mesh refinement needed to
obtain a good enough resolution of the deformed capsule. 

\subsection{Simple shear flow}\label{sec:linearshear}

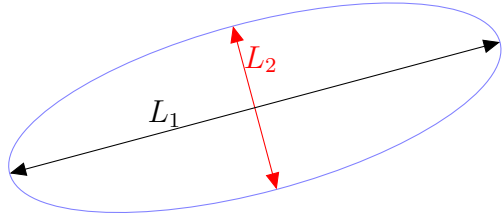
\begin{figure}
    \centering
    \begin{tikzpicture}[
    scale=0.8,
    >=triangle 45,
    mydeco/.style = {decoration = {markings, 
    mark = at position #1 with {\arrow{>}}}},
    baseline={(-2,0)}
]
    \draw[blue!50, rotate=-75] (0,0) ellipse (40pt and 120pt);

    \draw[<->, red, rotate=-75] (-40pt,0.0) -- (40pt,0.0); \draw (0.1,0.5) node[red,above]{$L_2$}; 
    \draw[<->, black, rotate=15] (-120pt,0.0) -- (120pt,0.0); \draw (-1.5,-0.4) node[black,above]{$L_1$}; 
   
\end{tikzpicture}
    \caption{Simple shear flow: steady state configuration of the deformed capsule, with $L_1$ and $L_2$ \revI{representing} the principal directions.}
    \label{fig:linear_shear_l1_l2}
\end{figure}

\begin{figure}
    \centering
	\subfloat[Results from TC1]{\includegraphics[width=.48\textwidth]{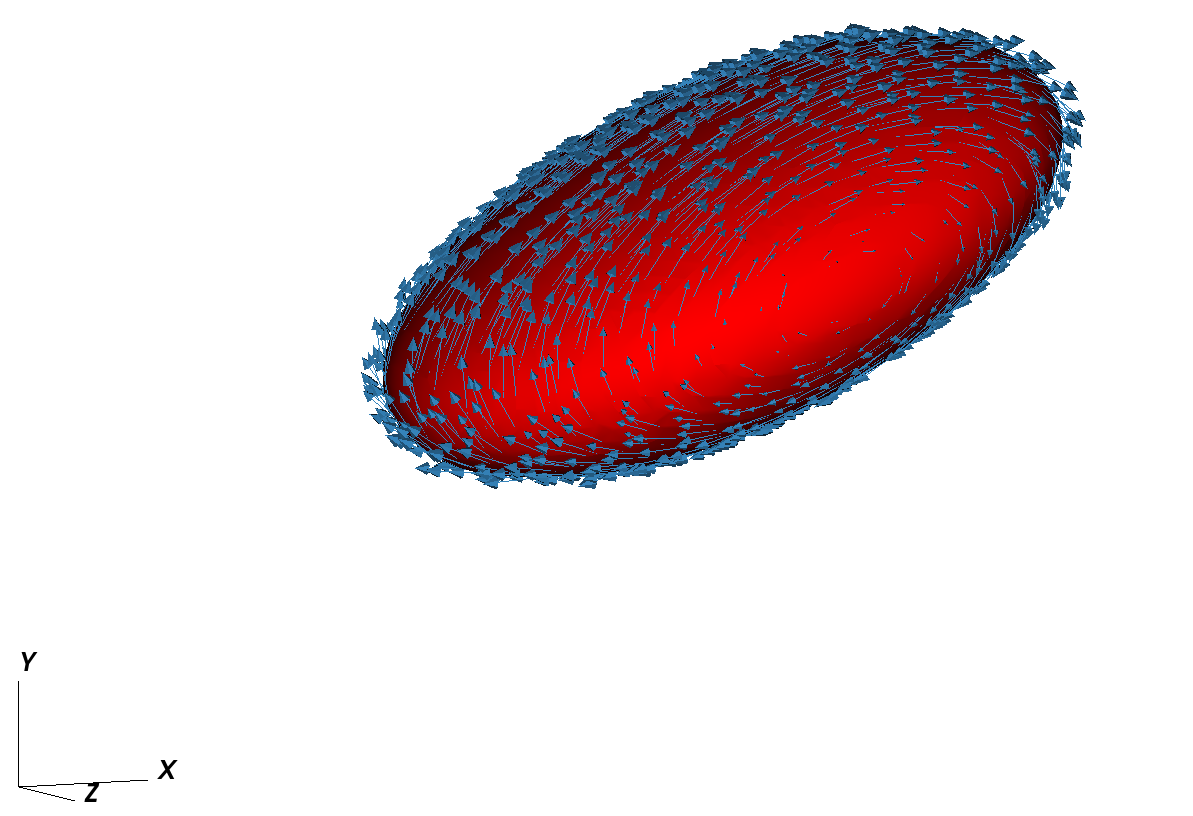}}\quad
	\subfloat[Results from TC4]{\includegraphics[width=.48\textwidth]{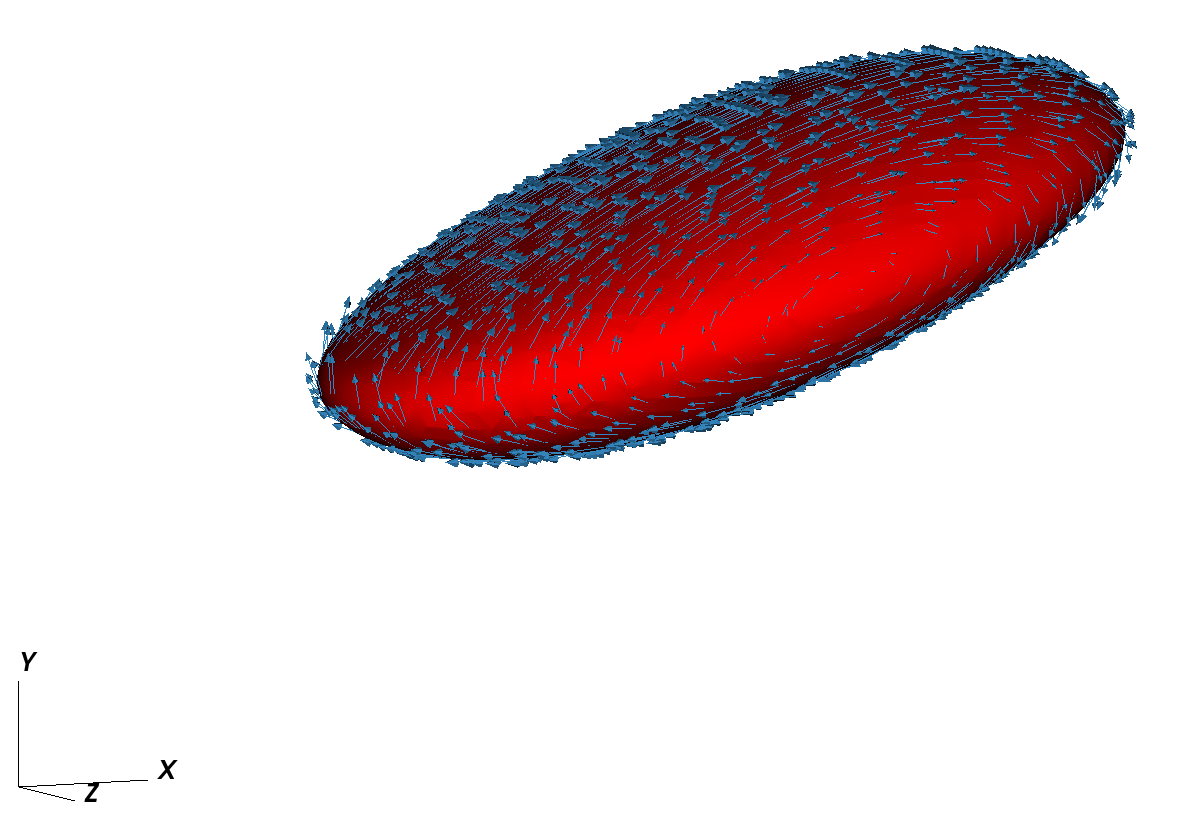}}
    \caption{Simple shear flow: three-dimensional visualization of the tank-treading motion at steady state.} 
    \label{fig:tanktreading}
\end{figure}

\begin{figure}
    \centering
    \subfloat[Data from TC1, TC2, TC3, TC4]{
     \begin{tikzpicture}[
       scale=0.95,
        >=triangle 45,
        mydeco/.style = {decoration = {markings, 
        mark = at position #1 with {\arrow{>}}}},
        baseline={(-2,0)}
        ]

        \begin{axis}[xmin=0,ymin=0,xlabel={$C_a$}, ylabel={$D_{12}$},
        legend style={at={(0.6,0.45)},anchor=north}]
            \addplot[mark =o,green] table[x index=0,y index=1] {Linear_shear/NH_Lac.txt};
            \addplot[mark =pentagon,blue] table[x index=0,y index=1] {Linear_shear/NH_Li.txt};
            \addplot[mark =+,red] table[x index=0,y index=1] {Linear_shear/NH_Walter.txt};
            \addplot[line width=1pt, mark =x,black,mark size=3pt] table[x index=0,y index=1] {Linear_shear/NH_Data.txt};
            
            \legend{Lac et al. ($2004$), Li et al. ($2008$), Walter et al. ($2010$), Present work};
        \end{axis}
    
\end{tikzpicture}}\qquad
    \subfloat[Data from TC5, TC6, TC7, TC8]{
     \begin{tikzpicture}[
       scale=0.95,
        >=triangle 45,
        mydeco/.style = {decoration = {markings, 
        mark = at position #1 with {\arrow{>}}}},
        baseline={(-2,0)}
        ]

        \begin{axis}[xmin=0,ymin=0, xlabel={$C_a$}, ylabel={$D_{12}$},
        legend style={at={(0.6,0.45)},anchor=north}]
            \addplot[mark =o,green] table[x index=0,y index=1] {Linear_shear/Skalak_Lac.txt};
            \addplot[mark =pentagon,blue] table[x index=0,y index=1] {Linear_shear/Skalak_Li.txt};
            \addplot[mark =+,red] table[x index=0,y index=1] {Linear_shear/Skalak_Walter.txt};
            \addplot[line width=1pt, mark =x,black,mark size=4pt] table[x index=0,y index=1] {Linear_shear/Skalak_Data.txt};
            
            \legend{Lac et al. ($2004$), Li et al. ($2008$), Walter et al. ($2010$), Present work};
        \end{axis}
    
\end{tikzpicture}}
    \caption{Simple shear flow: comparisons with respect to the dimensionless parameter $D_{12}$ with \revII{Neo--Hookean} (left) and Skalak (right) law types.}
    \label{fig:linear_shear_d12_comparison_NH_SK}
\end{figure}
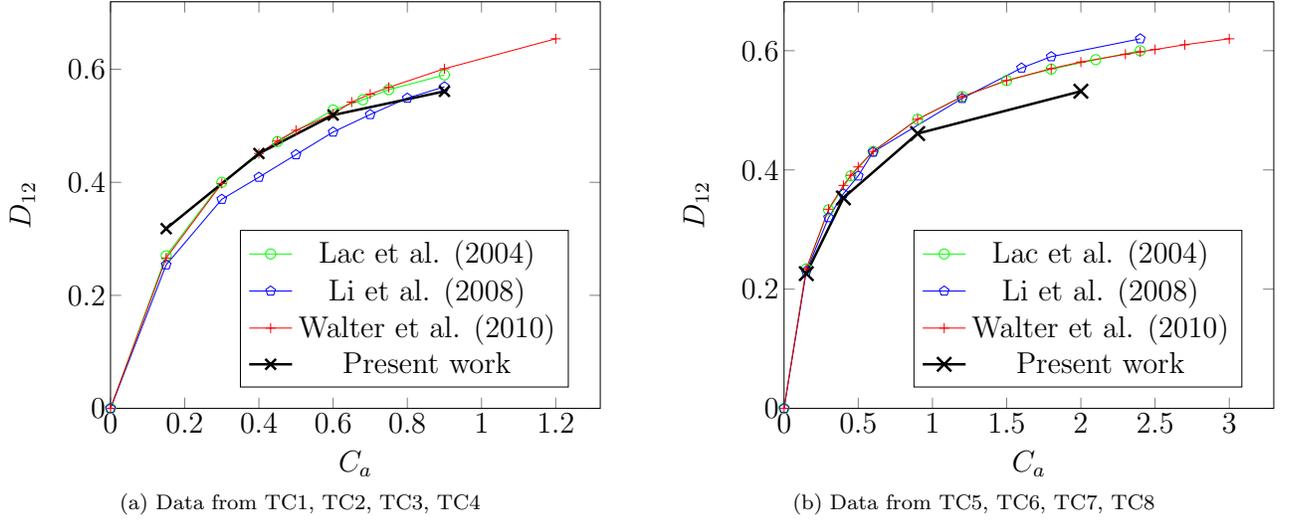

In this section, we present the numerical simulations for the simple shear flow test case. 
Thanks to the results available in the literature~\cite{Lac2004,Li2008,Walter2010}, this test case allows us to validate 
the proposed method with complex configurations, typical for capsule applications, 
arising when a membrane is immersed within a linear shear velocity field. 
The complexity comes from the fact that the steady state consists in a constant shear occurring on both the inside and the outside 
of the membrane. The numerical challenge of such test cases is related to evolution of the deformation vector $Y$ close to the interface. 
For this reason, it is mandatory for the long-time simulation of this test case to use both the Aslam extrapolation and the inner diffusion in order to capture 
properly the tank-treading motion (mentioned in Section~\ref{sec:aslamdiffusion}).\\
The simulation is setup by considering a thin membrane of radius $a_m$, modeled with several constitutive laws,
immersed within a viscous flow, characterized by the physical parameters in Table \ref{tab:param_linear_shear_fixed}, 
in the domain $[-4,4]\times[-2,2]\times[-2,2]$.
The dimensionless parameters for this configuration read
\begin{equation}
Re = \frac{\rho a_m^2 \dot\gamma}{\mu} = 0.0625, \qquad Ca = \frac{\mu \dot\gamma a_m}{G_s},
\end{equation}
where $Re$ is the Reynolds number, and $Ca$ is the capillary number, which is modified according to the configuration chosen. 
The initial conditions are given by Equation \eqref{eq:phisphere} to initialize the level-set of a sphere, and
\begin{equation}
u_0(x,y,z) = \begin{pmatrix} \dot\gamma y \\ 0 \\ 0 \end{pmatrix}, \qquad
Y_0(x,y,z) = \begin{pmatrix} x \\ y \\ z \end{pmatrix},
\end{equation}
to impose the linear shear velocity field and no pre-deformation on the initial configuration.
We impose Neumann boundary conditions along the $x$-axis, moving wall along the $y$-axis, and slip wall along the $z$-axis.\\
The computational domain is discretized using a uniform grid of $512 \times 256 \times 256=33\,554\,432$ cells.
The time step chosen for these simulations is set to $\Delta t=2\times 10^{-3}s$. 
\begin{table}[t]
    \centering
    \begin{tabular}{c|c|c|c}
         $\rho$ & $\mu$ & $a_m$ & shear rate $\dot \gamma$\\
         \hline
         1 & 4.0 & 0.5 & 1.0
    \end{tabular}
    \caption{Simple shear flow: physical parameters.}
    \label{tab:param_linear_shear_fixed}
\end{table}
Further details about the configurations simulated in this section are given in Table \ref{tab:param_linear_shear}, where
different constitutive law types and multiple values of the capillary number are tested. 
The constitutive laws employed for these simulations are the \revII{Neo--Hookean} \eqref{eq:law_neo_hookean} and the Skalak \eqref{eq:law_skalak} laws.
We remind here that, in general, the stretching modulus can be set as $K_s= 3G_s$.
\begin{table}[b]
    \centering
    \begin{tabular}{c||c|c|c|c}
         test case & $Ca$ &$G_s$ & $K_s$ & law type \\
         \hline
         TC1 & 0.15 & 13.33 & - & NH\\
         \hline
         TC2 & 0.4 & 5 & - & NH\\
         \hline
         TC3 & 0.6 & 3.33 & - & NH\\
         \hline
         TC4 & 0.9 & 2.22 & - & NH\\
         \hline
         TC5 & 0.15 & 13.33 & 40 & SK\\
         \hline
         TC6 & 0.4 & 5 & 15 & SK\\
         \hline
         TC7 & 0.9 & 2.22 & 6.66 & SK\\
         \hline
         TC8 & 2.0 & 1 & 3 & SK\\
    \end{tabular}
    \caption{Simple shear flow: stretching and shear moduli, $K_s$ and $G_s$, when varying the capillary number for the \revII{Neo--Hookean} (NK) and Skalak (SK) constitutive laws.}
    \label{tab:param_linear_shear}
\end{table}
In these contributions, the shape of the membrane is assumed to be an ellipsoid once the steady state is reached and, by taking the slice 
$z=0$ of this geometry, it is possible to introduce a dimensionless parameter depending on the characteristics of the ellipse,
\begin{equation}
D_{12} = \frac{L_1-L_2}{L_1+L_2}.
\end{equation}
This parameter, computed by taking the major and minor axes of the ellipse (see Figure \ref{fig:linear_shear_l1_l2}), has been mainly introduced for validation purposes. 
Here, the proposed approach is compared to the literature~\cite{Lac2004,Li2008,Walter2010}.
In Figure \ref{fig:tanktreading} we show two three-dimensional visualizations of the tank-treading motion for different capillary 
numbers by plotting the velocity vectors of the fluid on the membrane surface. 
Here we show the flow velocity field over the membrane surface to point out that even if the geometry of the capsule is no longer
changing, the structure dynamics is still evolving.
It is then clear from the visualization that the steady state configuration consists in a fixed but rotating membrane.
Figure \ref{fig:linear_shear_d12_comparison_NH_SK} presents the results of $D_{12}$ when varying the capillary number. 
The results presented in this section are in good agreement with those coming from the literature showing a reasonable increasing 
trend of $D_{12}$, when a larger capillary number is considered. 
\revI{The differences observed between our computed value of $D_{12}$ and those coming from the literature may be related 
to the different fluid modeling (Stokes vs.\ Navier--Stokes with low Reynolds) and numerical method, but also from the actual computation of the parameter $D_{12}$. As a matter of fact, 
when high capillary numbers are considered, the capsule shape takes very elongated form which no resembles a real ellipse
as also shown in~\cite{Walter2010}, which makes the evaluation of its semi-axes much more ambiguous.}
In agreement with the results in Figure \ref{fig:linear_shear_d12_comparison_NH_SK}, it is possible to observe also in 
Figure \ref{fig:tanktreading} that using smaller capillary numbers leads to a stiffer membrane that tends to elongate less.

\subsection{Capsule in a square-section channel}\label{sec:capsulesquare}

\begin{figure}
	\centering
	\subfloat[TC9 : slice on the $yz$-plane.]{\includegraphics[width=.30\textwidth]{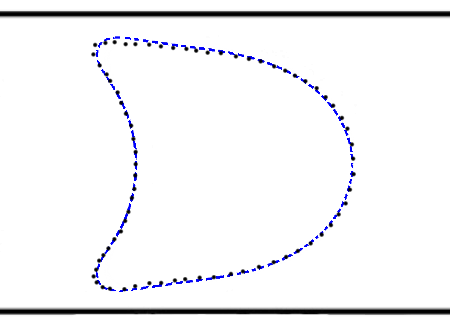}}\qquad
	\subfloat[TC10: slice on the $yz$-plane.]{\includegraphics[width=.30\textwidth]{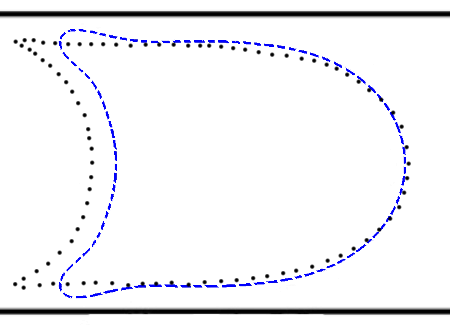}}\qquad\qquad
	\subfloat[TC9 : slice on the $xy$-plane.]{\includegraphics[width=.30\textwidth]{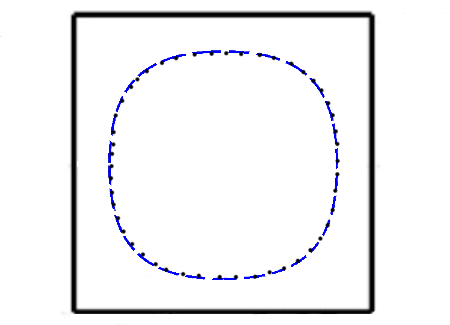}}\qquad
	\subfloat[TC10: slice on the $xy$-plane.]{\includegraphics[width=.30\textwidth]{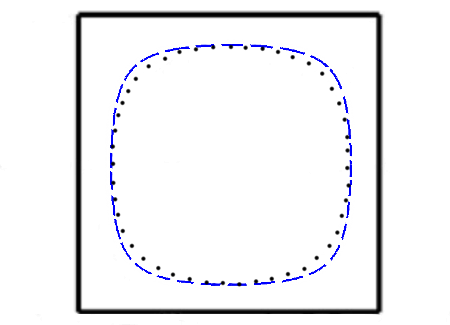}}
    \caption{Capsule in a square-section channel: the membrane profiles obtained by the proposed method with \revII{Neo--Hookean} constitutive law (dashed blue line) are compared to those presented in \cite{Hu2013} (black dotted line).}
    \label{fig:capsule_square_section_result_nh}
\end{figure}

\begin{figure}
	\centering
	\subfloat[TC11: slice on the $yz$-plane.]{\includegraphics[width=.3\textwidth]{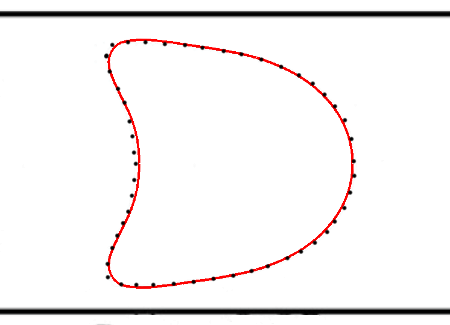}}\quad 
	\subfloat[TC12: slice on the $yz$-plane.]{\includegraphics[width=.3\textwidth]{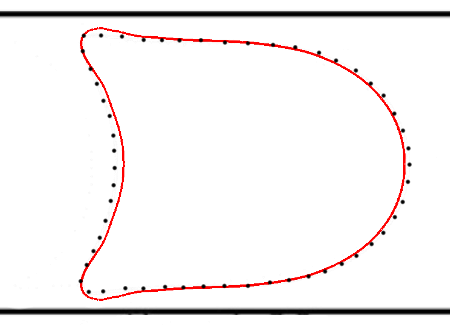}}\quad
	\subfloat[TC13: slice on the $yz$-plane.]{\includegraphics[width=.3\textwidth]{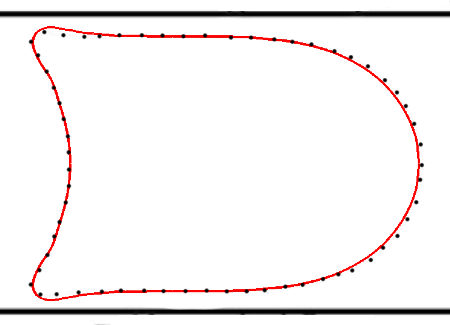}}\quad
	\subfloat[TC11: slice on the $xy$-plane.]{\includegraphics[width=.3\textwidth]{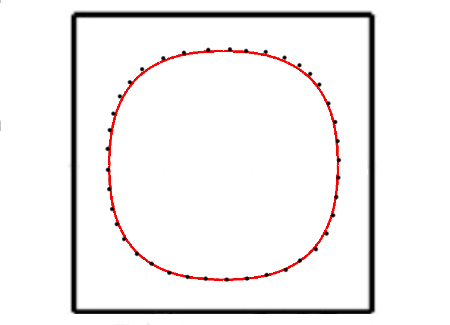}}\quad
	\subfloat[TC12: slice on the $xy$-plane.]{\includegraphics[width=.3\textwidth]{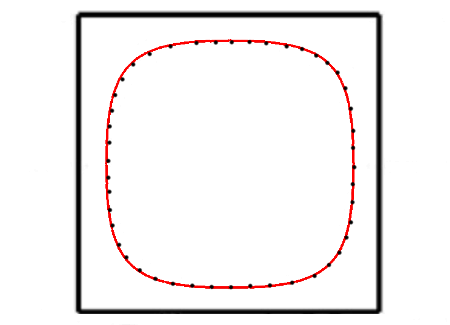}}\quad
	\subfloat[TC13: slice on the $xy$-plane.]{\includegraphics[width=.3\textwidth]{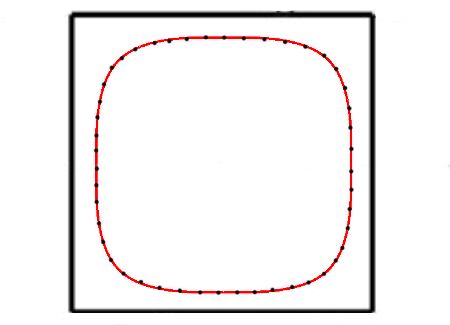}}
    \caption{Capsule in a square-section channel: the membrane profiles obtained by the proposed method with Skalak constitutive law (red bold line) are compared to those presented in \cite{Hu2013} (black dotted line).}
    \label{fig:capsule_square_section_result_sk}
\end{figure}

The test cases presented in this section have the goal of simulating capsules flowing in a square-section channel, 
even when the capsule size at rest is larger than the channel dimension. 
In these situations, an interaction between the membrane and the boundary layer occurs and the capsule deforms into a parachute shape.\\
The simulation is setup by considering a thin membrane, modeled with several constitutive laws,
immersed within a viscous flow in a channel of dimensions $[-1,1]\times[-0.5,0.5]\times[-0.5,0.5]$.
\revI{It should be noticed that the interaction between the membrane and the wall is not modeled with special techniques, but
by refining the computational mesh enough to have an accurate resolution of the boundary layer.}
The initial conditions for both the level-set field $\phi$ and deformation vector $Y$ are computed so that the capsule has a pre-deformed shape. 
This choice is related to the fact that, for some test cases, the capsule is assumed to be larger than the channel cross section.
In this case, the initial conditions read
\begin{equation}
\phi_0(x,y,z)=\sqrt{(xe^{-2t_0})^2 + (ye^{t_0})^2 + (ze^{t_0})^2} - a_m ,
\end{equation}
where $a_m$ is the radius of the relaxed sphere, and 
\begin{equation}
u_0(x,y,z) = \begin{pmatrix} V_p - V \\ 0 \\ 0\end{pmatrix}, \qquad Y_0(x,y,z) = \begin{pmatrix} xe^{-2t_0}\\ ye^{t_0}\\ ze^{t_0} \end{pmatrix},
\end{equation}
where $t_0$ is the pre-deformation time, $V$ is the mean capsule velocity, and 
$V_p$ is a two-dimensional Poiseuille on a square section \cite{Pozrikidis1997} such that,
\begin{equation*}\label{eq:poiseuille2D}
V_p = \frac{\pi V}{2\beta} \sum\limits_{i=0}^{\infty}\left( \frac{1}{(2i+1)^3} - \frac{\cosh\left(\frac{(2i+1) \pi z}{\ell}\right)}{(2i+1)^3 \cosh\left(\frac{(2i+1)\pi}{2}\right)}\right)\sin\left((2i-1)\pi\left(\frac{y}{\ell}+\frac{1}{2}\right)\right),
\end{equation*}
where
\begin{equation*}
\beta = \frac{\pi^4}{96} - \frac2\pi \sum\limits_{i=0}^{\infty}  \frac{\tan\left( \frac{(2i+1)\pi}{2} \right)}{(2i+1)^5 } .
\end{equation*}
The equations on $\phi_0$ and $Y_0$ represent those of a deformed sphere into an ellipsoid elongated, by a virtual 
divergence-free velocity field, along the $x$-axis and compressed along the $y$ and $z$ axes. 
We impose inlet and Neumann conditions along the $x$-axis, and moving wall on the other boundaries. 
\revII{It should be noticed that these simulations are set up on a very short channel to save the computational resources 
needed to perform the tests on a much longer one, which allows the capsule to reach the steady state position. 
Similarly to what happens in a wind tunnel, the fluid is moving, while the structure stays at its place. 
However, since the capsule is deforming, its mean axial velocity $V$ is added onto the inlet condition on the right, 
but also on the moving walls such that the capsule does not exit the domain. In general, the same results can be obtained
by making a classical setup in a longer channel as shown in Section \ref{sec:simulationbulk}.}
Multiple configurations are studied by varying the constitutive laws used to model the elastic properties of the membrane and the ratio $a/\ell$, where $2\ell=1$ represents the channel's size. 
Further details about the numerical parameters chosen in this section are given in Table \ref{tab:param_square_cross}. 
All simulations are setup by considering the following dimensionless parameters,
\begin{equation}
Re = \frac{\rho a_m V}{\mu} = 0.05, \qquad Ca = \frac{\mu V}{G_s}=0.1,
\end{equation}
where $\rho$ is set to one. 
The goal of this benchmark is analyzing the deformation of the parachute shape with respect to the influence of the constitutive law and the ratio $a_m/\ell$.\\
The computational domain is discretized using a uniform grid of $128\times64\times64=524\,288$ cells and
the time step is set to $\Delta t = 10^{-3} s$.
\begin{table}
    \centering
    \begin{tabular}{c||c|c|c|c|c|c}
         test case & $a_m/\ell$ & $t_0$ & $\mu$ &$G_s$ & $K_s$ & law type \\
         \hline
         TC9  & 0.85 & 0.0 & 8.5 & 85 & - & NH\\
         \hline
         TC10 & 1.00 & 0.1 & 10 & 100 & - & NH\\
         \hline
         TC11 & 0.85 & 0.0 & 8.5 & 85 & 255 & SK\\
         \hline
         TC12 & 1.00 & 0.1 & 10 & 100 & 300 & SK\\
         \hline
         TC13 & 1.10 & 0.2 & 11 & 110 & 330 & SK\\
    \end{tabular}
    \caption{Capsule in a square-section channel: stretching and shear moduli, $K_s$ and $G_s$ ,
when varying the ratio $a_m/\ell$ for the \revII{Neo--Hookean} (NK)
and Skalak (SK) constitutive laws.}
    \label{tab:param_square_cross}
\end{table}
In Figure \ref{fig:capsule_square_section_result_nh} we present the numerical results obtained when modeling the elastic membrane with \revII{Neo--Hookean} constitutive law and varying the size of the immersed capsule. 
Figure \ref{fig:capsule_square_section_result_sk} shows similar numerical results obtained by using the Skalak law, also 
with a capsule larger than the channel (TC14) to assess our method for highly compressed capsules.
Due to the small size of the channel, the capsule is highly influenced by the higher velocity profile occurring in the middle of the channel that brings the structure to take an elongated shape that resembles that of a phantom.
Increasing the size of the capsule leads to an even more elongated parachute, whose shape even features corners close to the boundary layer (see Figure \ref{fig:capsule_square_section_result_nh}b). 
\revIII{The majority of capsule profiles presented in Figures \ref{fig:capsule_square_section_result_nh} and 
\ref{fig:capsule_square_section_result_sk} show a good agreement with the numerical results presented in \cite{Hu2013}.
The shape difference experienced for the Neo--Hookean law shown in Figure \ref{fig:capsule_square_section_result_nh}b may be related
to the differences in the fluid model (Stokes in that case) and interface treatment used to solve the problem. 
In particular, the fact that in \cite{Hu2013} a Lagrangian triangulation is used to follow the capsule deformation might 
be a reason to the better capture of the sharp corners experienced behind the capsule. 
However, in \cite{Hu2013}, the authors perform an experimental study where such sharp corners seem not to be 
present for real capsules of the size of the channel. Similar shapes seem only to be experienced by bigger capsules in real experiments, 
as also predicted in Figure~\ref{fig:capsule_square_section_result_sk}c. }


\section{Numerical simulations of a capsule with an internal nucleus with complex geometries}\label{sec:simulationbulk}

In this section, we present the numerical simulations of a highly compressed capsule travelling within a square-section channel until it opens to a much wider space bringing the capsule to feature a relaxation phenomenon. 
A novelty with respect to previous works is the introduction of a solid bulk structure within the thin membrane. 
\revI{In particular, we perform a wide set of simulations performed by varying the mechanical properties and size on the solid nucleus,
and provide all characteristic curves and deformed capsule shapes such that both numericists and experimentalists may use them for comparisons.}

\begin{figure}
    \centering
    \includegraphics[width=.85\textwidth]{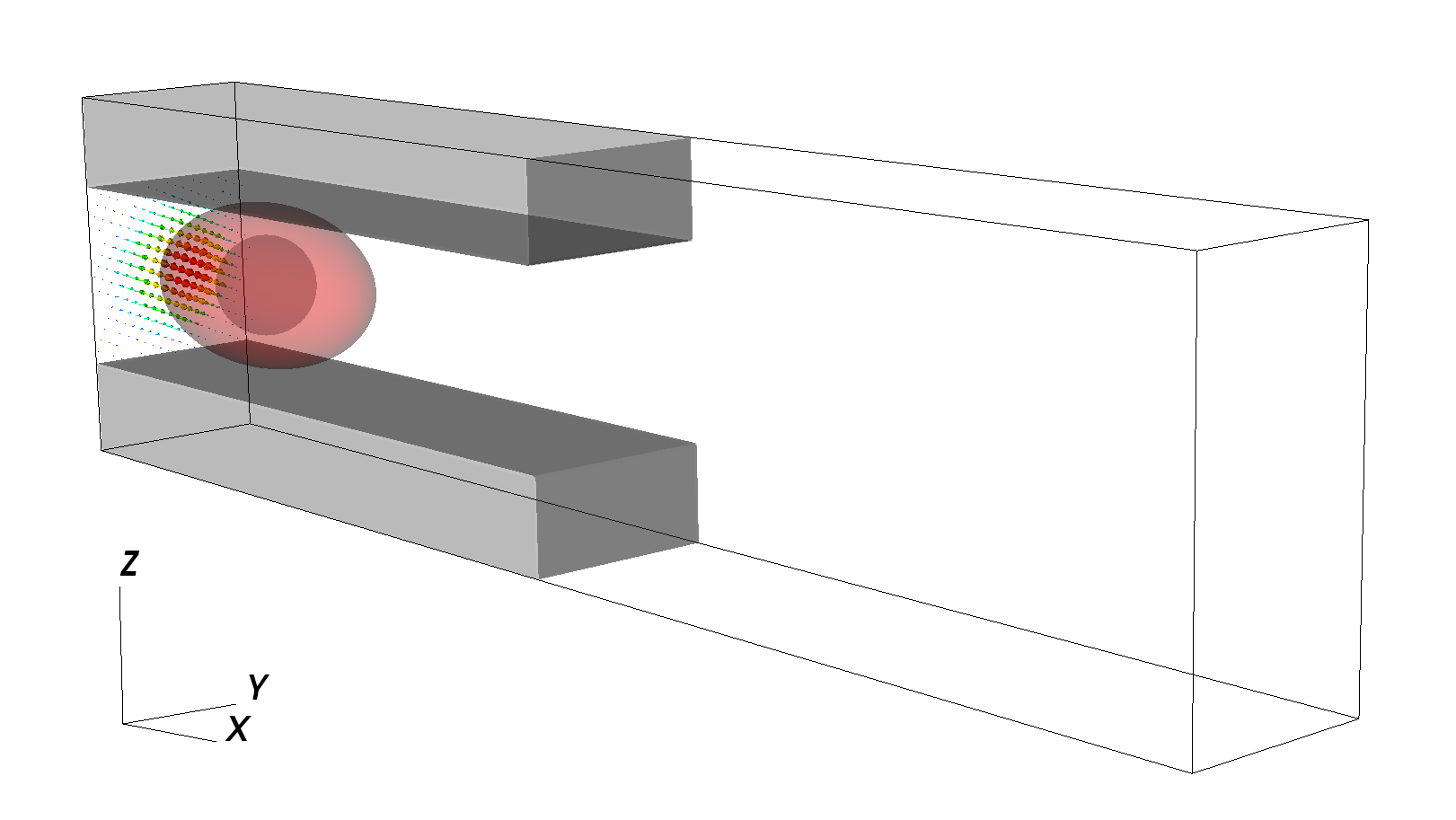}
    \caption{Capsule relaxation phenomenon: initial configuration of a membrane with an internal solid.}
    \label{fig:relaxation_init}
\end{figure}
The domain is divided in two part: the first one being a square-section channel $[-1, 3]\times[-0.5, 0.5]\times[-0.5, 0.5]$ where the compressed capsule is initialized and the second one being the wider part $[3, 7]\times[-1, 1]\times[-0.5, 0.5]$ where the relaxation happens (see Figure \ref{fig:relaxation_init}). 
The simulation is setup by considering a thin membrane, of radius $a_m$, with an internal bulk of radius $a_b$, immersed within a
viscous flow. 
The internal square-section channel is treated as an immersed boundary by imposing wall boundary conditions through a classical 
penalization approach.

\begin{figure}
\centering
\subfloat[Without internal bulk]{\includegraphics[width=.8\textwidth]{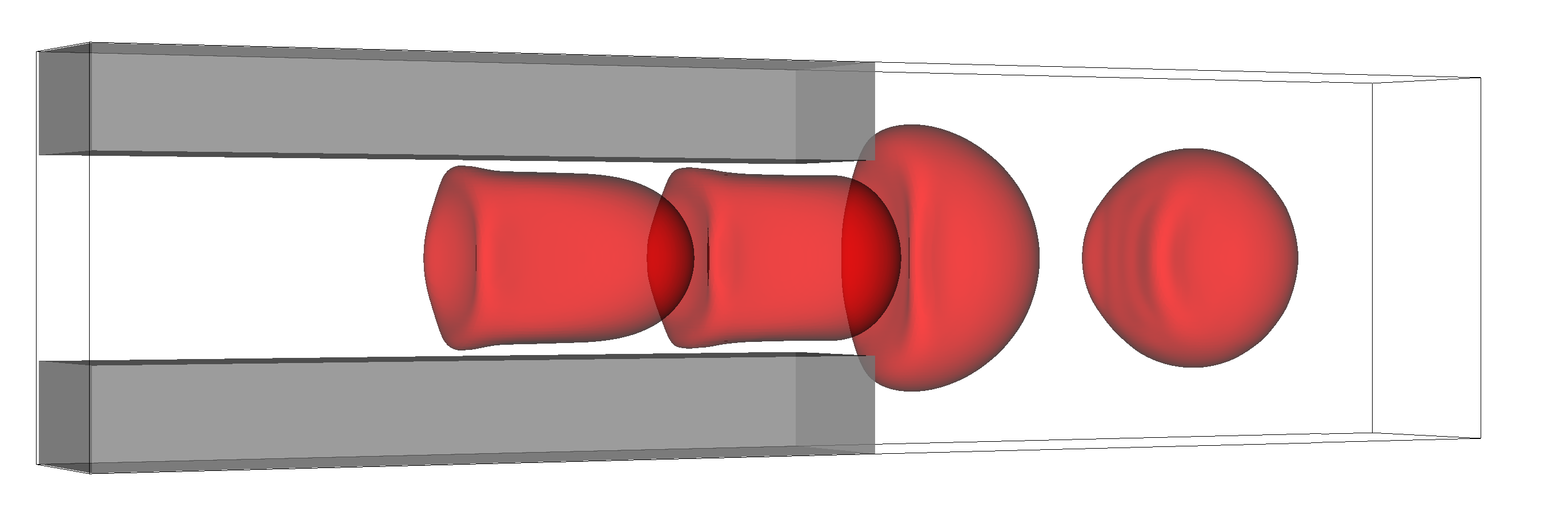}}\\
\subfloat[With internal bulk]   {\includegraphics[width=.8\textwidth]{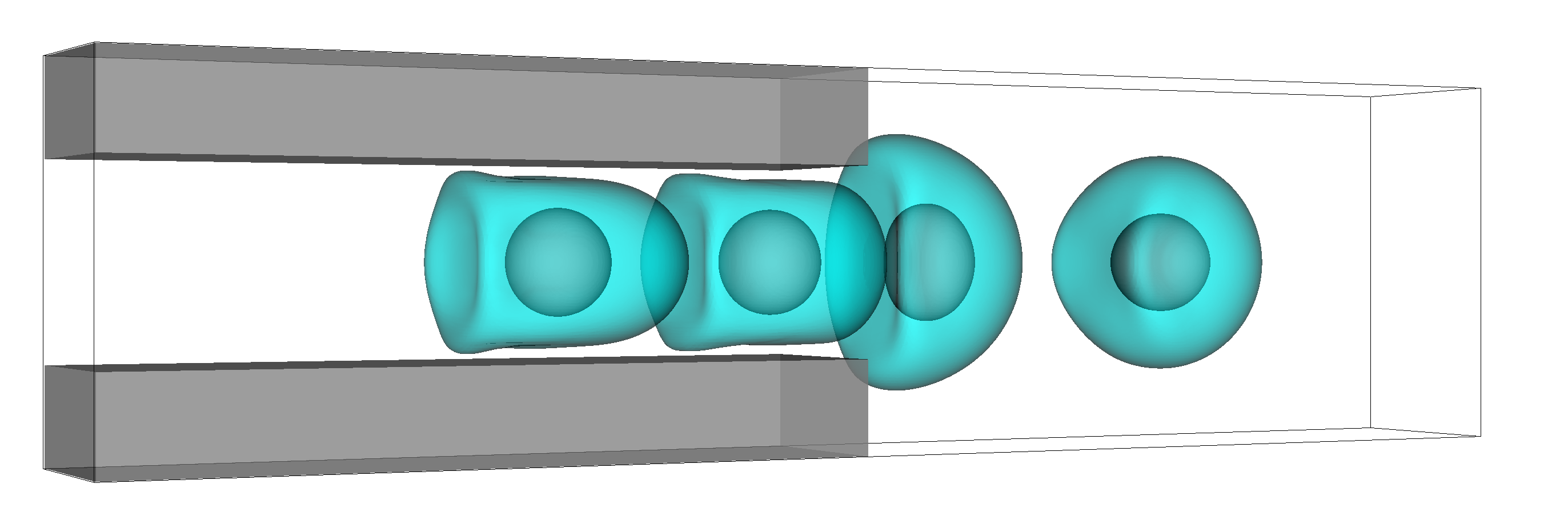}}
\caption{Capsule relaxation phenomenon: evolution of the capsule profile at different times.}
\label{fig:relaxation_surface_TC14}
\end{figure}

\begin{figure}
\centering
\begin{tikzpicture}
\node at (0,0){\includegraphics[width=0.45\textwidth]{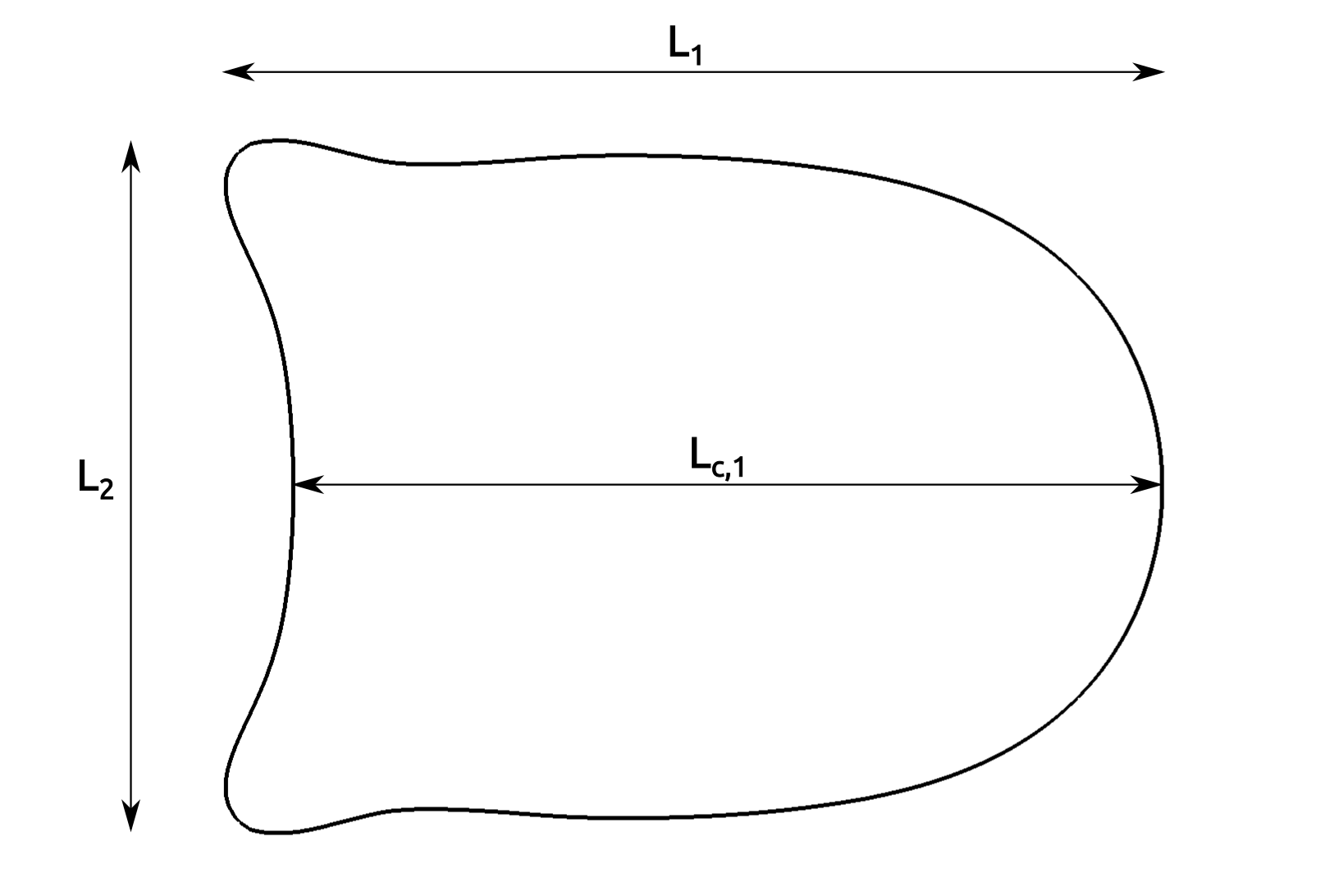}};
\node[rectangle, fill = white] (r) at (0.35,-0.05) {$L_3$};
\node[rectangle, fill = white] (r) at (-3.35,-0.20) {$L_2$};
\node[rectangle, fill = white] (r) at (0.2,2.40) {$L_1$};
\end{tikzpicture}
\caption{Capsule relaxation phenomenon: $L_1$, $L_2$, $L_3$ are the characteristic length of deformed capsule.}
\label{fig:relaxation_scheme_lengths}
\end{figure}
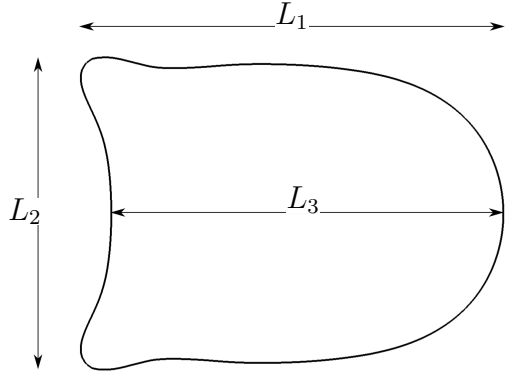

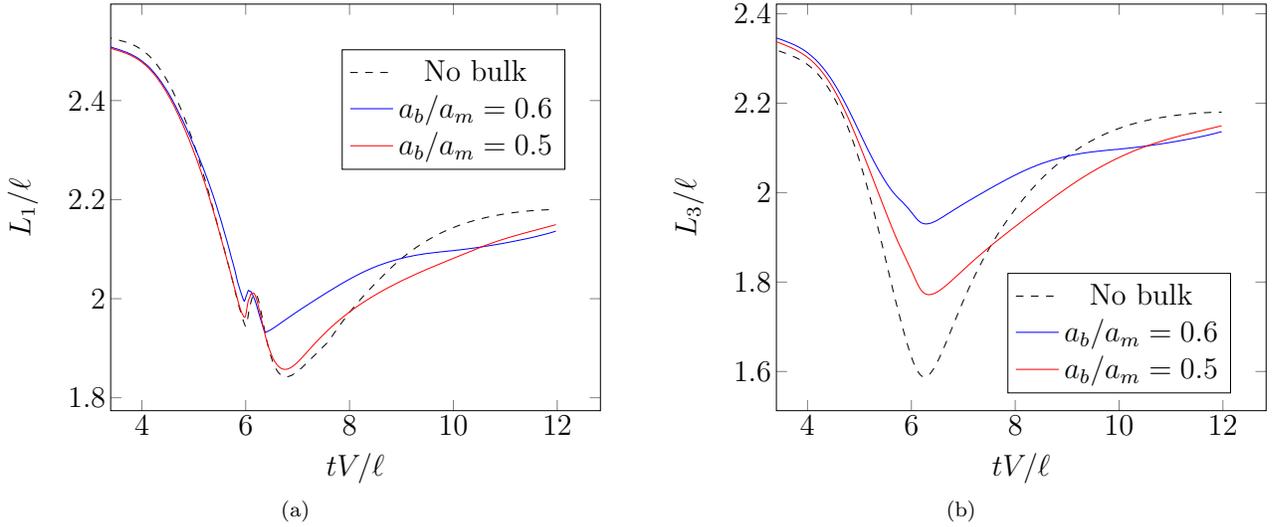
\begin{figure}
\centering
\subfloat[]{
\begin{tikzpicture}[
    scale=0.95,
    >=triangle 45,
    mydeco/.style = {decoration = {markings,
    mark = at position #1 with {\arrow{>}}}},
    baseline={(-2,0)}
    ]

    \begin{axis}[xlabel={$tV/\ell$}, ylabel={$L_{1}/\ell$},
    legend style={at={(0.7,0.9)},anchor=north},xmin=3.4]
    \addplot[black, dashed, smooth] table[x expr=\thisrowno{0}*2, y expr=\thisrowno{1}*2] {Capsule_square_relaxation/data/Relaxation_Lx_Ly_membrane.txt};
    \addplot[blue,smooth] table[x expr=\thisrowno{0}*2, y expr=\thisrowno{1}*2] {Capsule_square_relaxation/data/Relaxation_Lx_Ly_mixed_R1R2-06_Chi-20.txt};
    \addplot[red,smooth] table[x expr=\thisrowno{0}*2, y expr=\thisrowno{1}*2] {Capsule_square_relaxation/data/Relaxation_Lx_Ly_mixed_R1R2-05_Chi-20.txt};
    \legend{No bulk, $a_b/a_m=0.6$, $a_b/a_m=0.5$}
    \end{axis}
\end{tikzpicture}}\qquad
\subfloat[]{
\begin{tikzpicture}[
    scale=0.95,
    >=triangle 45,
    mydeco/.style = {decoration = {markings,
    mark = at position #1 with {\arrow{>}}}},
    baseline={(-2,0)}
    ]

    \begin{axis}[xlabel={$tV/\ell$}, ylabel={$L_3/\ell$},
    legend style={at={(0.7,0.35)},anchor=north},xmin=3.4]
    \addplot[black, dashed, smooth] table[x expr=\thisrowno{0}*2, y expr=\thisrowno{3}*2] {Capsule_square_relaxation/data/Relaxation_Lx_Ly_membrane.txt};
    \addplot[blue,smooth] table[x expr=\thisrowno{0}*2, y expr=\thisrowno{3}*2] {Capsule_square_relaxation/data/Relaxation_Lx_Ly_mixed_R1R2-06_Chi-20.txt};
    \addplot[red,smooth] table[x expr=\thisrowno{0}*2, y expr=\thisrowno{3}*2] {Capsule_square_relaxation/data/Relaxation_Lx_Ly_mixed_R1R2-05_Chi-20.txt};
    \legend{No bulk, $a_b/a_m=0.6$, $a_b/a_m=0.5$}
    \end{axis}
\end{tikzpicture}}
\caption{Capsule relaxation phenomenon: time evolution of $L_{1}/\ell$ and $L_3/\ell$ when varying the size of the internal bulk, while keeping a constant elastic modulus $\chi=2200$. The numerical results of TC14 are depicted using a dashed line, TC15 in blue, and TC16 in red.}
\label{fig:square_relaxation_L1Lc1_fixedchi}
\end{figure}

\begin{figure}
    \centering
    \begin{tikzpicture}[
       scale=1.2,
        >=triangle 45,
        mydeco/.style = {decoration = {markings,
        mark = at position #1 with {\arrow{>}}}},
        baseline={(-2,0)}
        ]

        \begin{axis}[xlabel={$tV/\ell$}, ylabel={$L_{2}/\ell$},
        legend style={at={(0.7,0.35)},anchor=north},xmin=3.4]
            \addplot[black, dashed, smooth] table[x expr=\thisrowno{0}*2, y expr=\thisrowno{2}*2] {Capsule_square_relaxation/data/Relaxation_Lx_Ly_membrane.txt};
            \addplot[blue,smooth] table[x expr=\thisrowno{0}*2, y expr=\thisrowno{2}*2] {Capsule_square_relaxation/data/Relaxation_Lx_Ly_mixed_R1R2-06_Chi-20.txt};
            \addplot[red,smooth] table[x expr=\thisrowno{0}*2, y expr=\thisrowno{2}*2] {Capsule_square_relaxation/data/Relaxation_Lx_Ly_mixed_R1R2-05_Chi-20.txt};
            \legend{No elastic volume, $a_b/a=0.6$, $a_b/a=0.5$}
        \end{axis}

    \end{tikzpicture}
    \caption{Temporal evolution of the length $L_2/\ell$ for three test case: no elastic volume ($TC15$), $a_b/a=0.6$ ($TC16$) and $a_b/a=0.5$ ($TC17$). The elastic volume coefficient is $\chi=2200$.}
    \label{fig:square_relaxation_L2_fixedchi}
\end{figure}
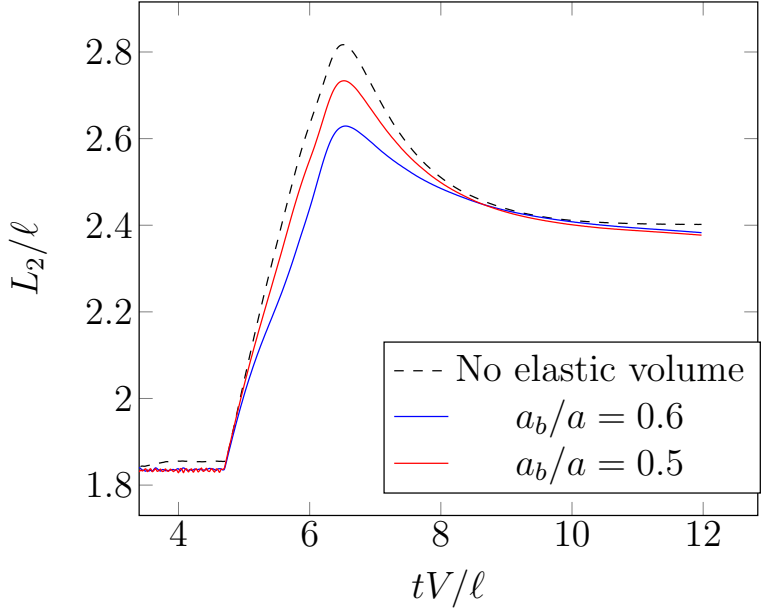

The dimensionless parameters used for this simulation are computed considering,
\begin{equation}
Re = \frac{\rho a_m V}{\mu} = 0.05, \qquad Ca = \frac{\mu V}{G_s}=0.1,
\end{equation}
and are shown in Table~\ref{tab:param_relaxation_fixed}, where $2\ell$ represents the channel dimension with $\ell=0.5$.
The initial condition on the velocity is given by the two-dimensional Poiseuille on a square-section given in Section \ref{sec:capsulesquare}. 
The level-set $\phi_m$ and deformation vector $Y_m$ are initialized as
\begin{equation}
\phi_{m,0}(x,y,z)=\sqrt{(xe^{-2t_0})^2 + (ye^{t_0})^2 + (ze^{t_0})^2} - a_m , \qquad Y_{m,0}(x,y,z) = \begin{pmatrix} xe^{-2t_0}\\ ye^{t_0}\\ ze^{t_0} \end{pmatrix},
\end{equation}
where $t_0=0.2$. The level-set $\phi_b$ and deformation vector $Y_b$ for the elastic bulk are initialized as a sphere with no pre-deformation,
\begin{equation}
\phi_{b,0}(x,y,z)=\sqrt{x^2 + y^2 + z^2} - a_b ,\qquad Y_{b,0}(x,y,z) = \begin{pmatrix} x\\ y\\ z \end{pmatrix}.
\end{equation}
As mentioned above, several simulations have been performed by varying the size $a_b$ and the elastic modulus $\chi$ of the internal bulk. 
Further details about the setup of both $a_b$ and $\chi$ are given in Table \ref{tab:param_relaxation}.\\
The computational domain is discretized using a uniform grid of $512\times128\times64=4\,194\,304$ cells, and the time step chosen is $\Delta t=5\times10^{-4}s$.
\begin{table}
    \centering
    \begin{tabular}{c|c|c|c|c|c|c|c}
         $\rho$ & $\mu$ & $V$ & $a_m$ & $G_s$ & $K_s$ & law type \\
         \hline
         1 & 11 & 1 & 0.55 & 110 & 330 & SK
    \end{tabular}
    \caption{Capsule relaxation phenomenon: simulation setup for the membrane.}
    \label{tab:param_relaxation_fixed}
\end{table}  
\begin{table}[H]
    \centering
    \begin{tabular}{c||c|c}
         test case & $\chi$ & $a_b/a_m$ \\
         \hline
         TC14 & - & No bulk\\
         \hline
         TC15 & 2200 & 0.6\\
         \hline
         TC16 & 2200 & 0.5\\
         \hline
         TC17 & 1100 & 0.5\\
         \hline
         TC18 & 220 & 0.5\\
         \hline
         TC19 & 110 & 0.5\\
         \hline
         TC20 & 2200 & 0.4\\
         \hline
         TC21 & 2200 & 0.2\\
    \end{tabular}
    \caption{Capsule relaxation phenomenon: simulation setup for different configurations of the internal bulk.}
    \label{tab:param_relaxation}
\end{table}
In Figure \ref{fig:relaxation_surface_TC14}, we plotted the evolution of the membrane profile, with and without an internal bulk. 
At the beginning, as presented in Section \ref{sec:capsulesquare}, when flowing within the square-section channel, 
the compressed capsule takes a phantom shape. 
Then, once the capsule is at the end of the channel, it starts relaxing and modifying its shape until it becomes again similar to a sphere.
%
%
To give more insights on the simulations, in Figure~\ref{fig:relaxation_scheme_lengths} we define three characteristic lengths to study the evolution of the parachute during the relaxation phenomenon.
\revI{Figure \ref{fig:square_relaxation_L1Lc1_fixedchi} shows the evolution in non-dimensional time $tV/\ell$ of $L_1/\ell$ and $L_3/\ell$ 
when varying the bulk's size $a_b/a_m$. For these cases, we consider a constant elastic modulus $\chi=2200$ (TC14, TC15, and TC16 in Table \ref{tab:param_relaxation}).} 
As expected, in all cases $L_1/\ell$ suddenly decreases and features a small jump when the capsule is exiting the channel. 
After that, the relaxation process begins and $L_1/\ell$ increases again. 
Figure \ref{fig:square_relaxation_L1Lc1_fixedchi} shows that the presence of an elastic bulk slightly influences the evolution of $L_1/\ell$.
Similar trends are observed for $L_3/\ell$, however it should be noticed that 
in this case the presence of an elastic bulk introduces remarkable changes in the 
parachute shape, which appears much more elongated for larger bulks.
In particular, it is observed that when increasing the bulk's size, the minimum of $L_3/\ell$ increases too. 
Although the dynamics of the relaxation process is highly influenced by the internal bulk, it is interesting to observe that the minimum value of $L_3/\ell$ always occurs around $tV/\ell=6.2$.
In Figure~\ref{fig:square_relaxation_L2_fixedchi} we plot the evolution of $L_2/\ell$ where the strong correlation with $L_1/\ell$ is observed: when $L_1/\ell$ decreases, $L_2/\ell$ increases and vice versa. 

\begin{figure}
    \centering
    \begin{tikzpicture}[
       scale=1.2,
        >=triangle 45,
        mydeco/.style = {decoration = {markings,
        mark = at position #1 with {\arrow{>}}}},
        baseline={(-2,0)}
        ]

        \begin{axis}[xlabel={$tV/\ell$}, ylabel={$L_3/\ell$},
        legend style={at={(0.75,0.45)},anchor=north},xmin=3.4]
            \addplot[black, dashed, smooth] table[x expr=\thisrowno{0}*2, y expr=\thisrowno{3}*2] {Capsule_square_relaxation/data/Relaxation_Lx_Ly_membrane.txt};
            \addplot[blue,smooth] table[x expr=\thisrowno{0}*2, y expr=\thisrowno{3}*2] {Capsule_square_relaxation/data/Relaxation_Lx_Ly_mixed_R1R2-05_Chi-20.txt};
            \addplot[red,smooth] table[x expr=\thisrowno{0}*2, y expr=\thisrowno{3}*2] {Capsule_square_relaxation/data/Relaxation_Lx_Ly_mixed_R1R2-05_Chi-10.txt};
            \addplot[orange,smooth] table[x expr=\thisrowno{0}*2, y expr=\thisrowno{3}*2] {Capsule_square_relaxation/data/Relaxation_Lx_Ly_mixed_R1R2-05_Chi-2.txt};
            \addplot[green,smooth] table[x expr=\thisrowno{0}*2, y expr=\thisrowno{3}*2] {Capsule_square_relaxation/data/Relaxation_Lx_Ly_mixed_R1R2-05_Chi-1.txt};
            \legend{No bulk, $\chi=2200$, $\chi=1100$, $\chi=220$, $\chi=110$}
        \end{axis}

    \end{tikzpicture}
\caption{Capsule relaxation phenomenon: time evolution of $L_3/\ell$ when varying the elastic modulus of the internal bulk, while keeping a constant radius ratio $a_b/a=0.5$. The numerical results of TC14 are depicted using a dashed line, TC16 in blue, TC17 in red, TC18 in orange, and TC19 in green.}
    \label{fig:square_relaxation_Lc1_fixedab}
\end{figure}
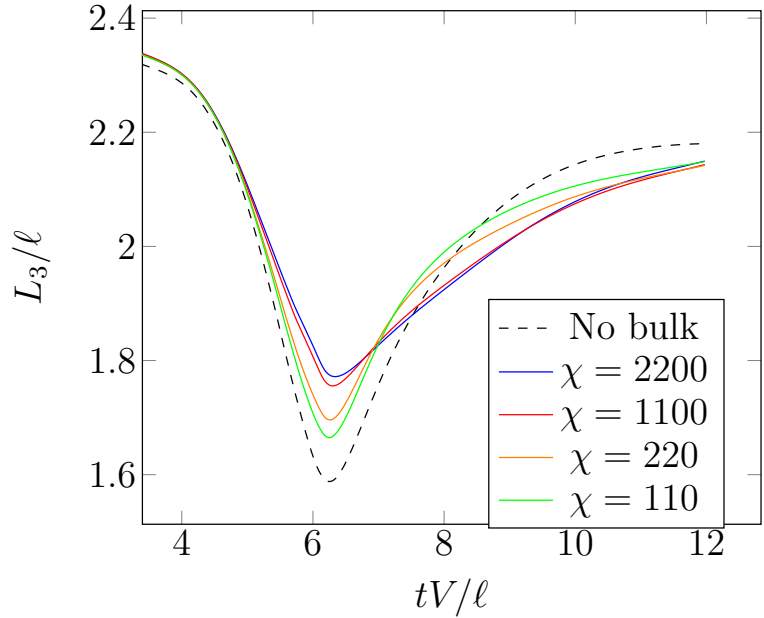

\begin{figure}
\centering
\subfloat[$tV/\ell=5.75$]{\includegraphics[width=.4\textwidth]{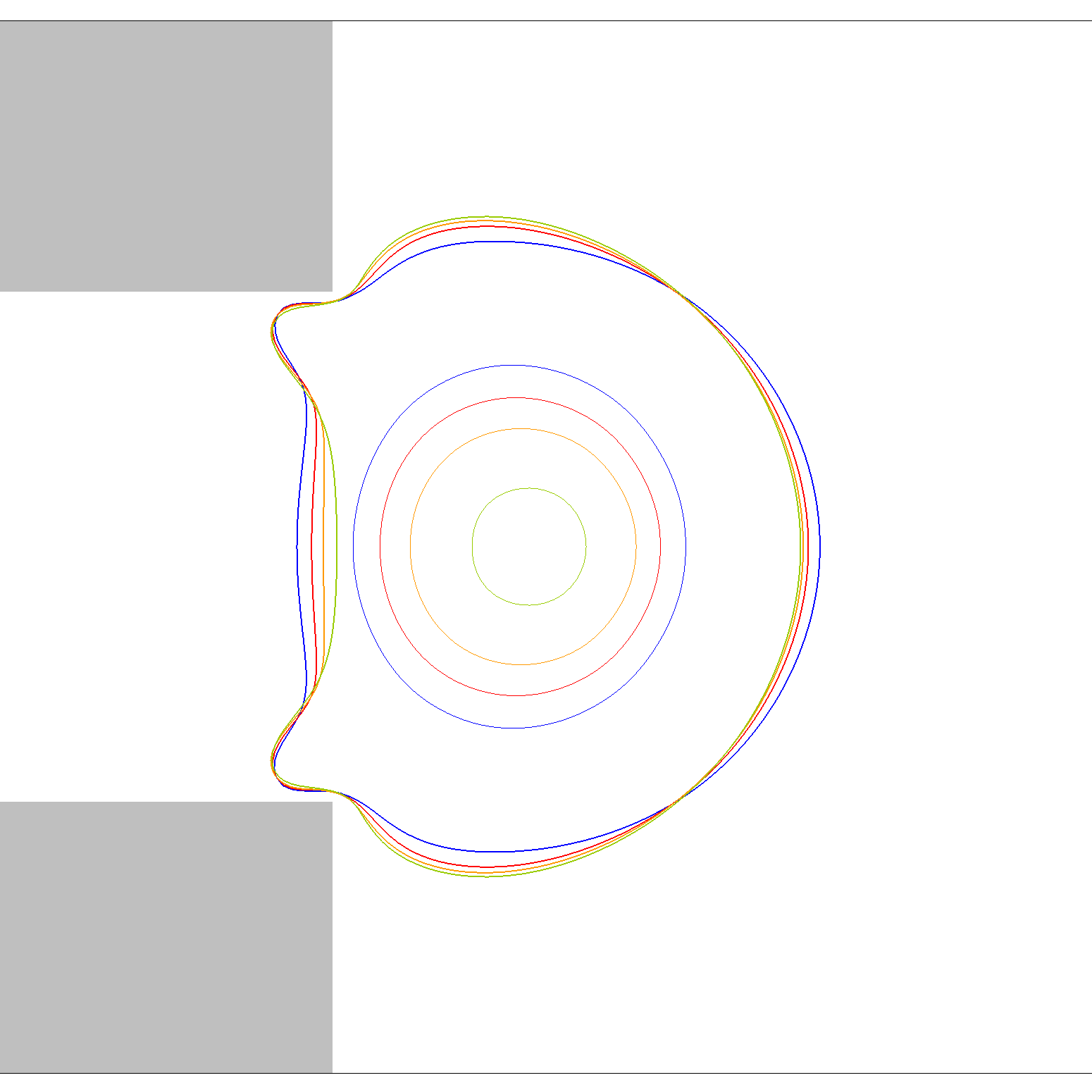}}\qquad\qquad
\subfloat[$tV/\ell=6.25$]{\includegraphics[width=.4\textwidth]{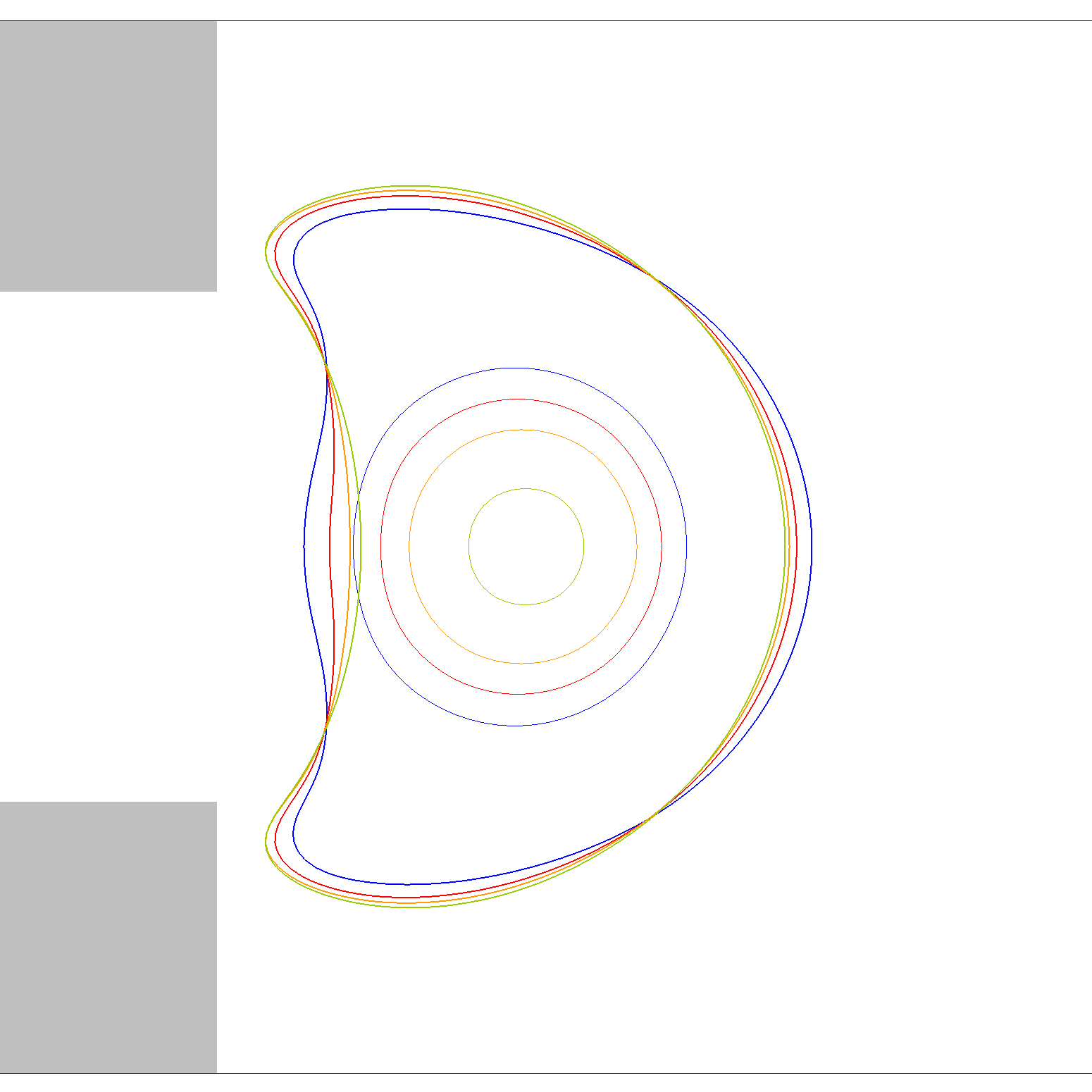}}
\caption{Capsule relaxation phenomenon: time evolution of the membrane and bulk profiles during the relaxation process when varying  the size of the internal bulk, while keeping a constant elastic modulus $\chi=2200$. The numerical results of TC15 are depicted in blue, TC16 in red, TC20 in orange, and TC21 in green. The walls of the square-section channel are represented in grey.}
\label{fig:relaxation_allChi20}
\end{figure}

\begin{figure}
\centering
\subfloat[$tV/\ell=5.75$]{\includegraphics[width=.4\textwidth]{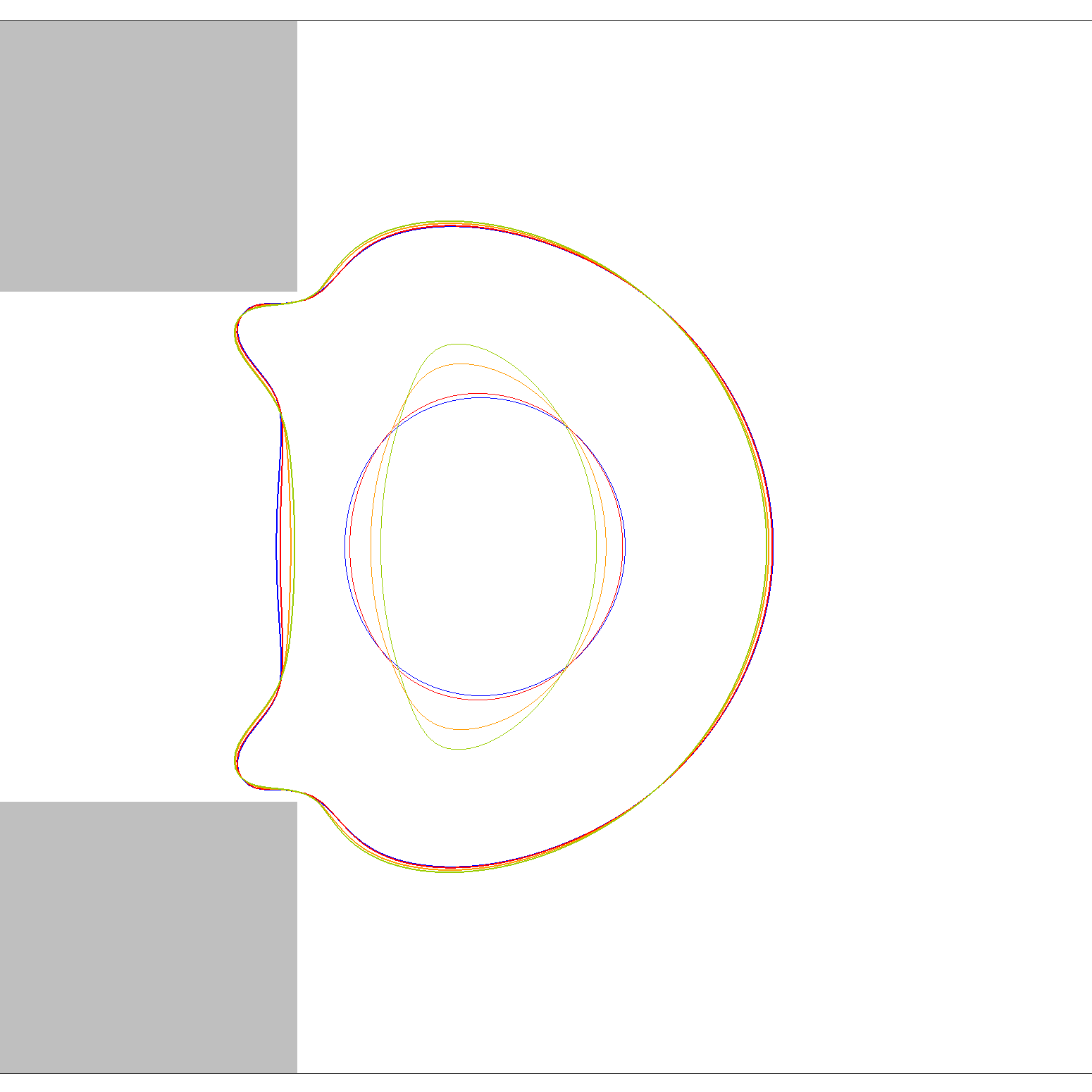}}\qquad\qquad
\subfloat[$tV/\ell=6.25$]{\includegraphics[width=.4\textwidth]{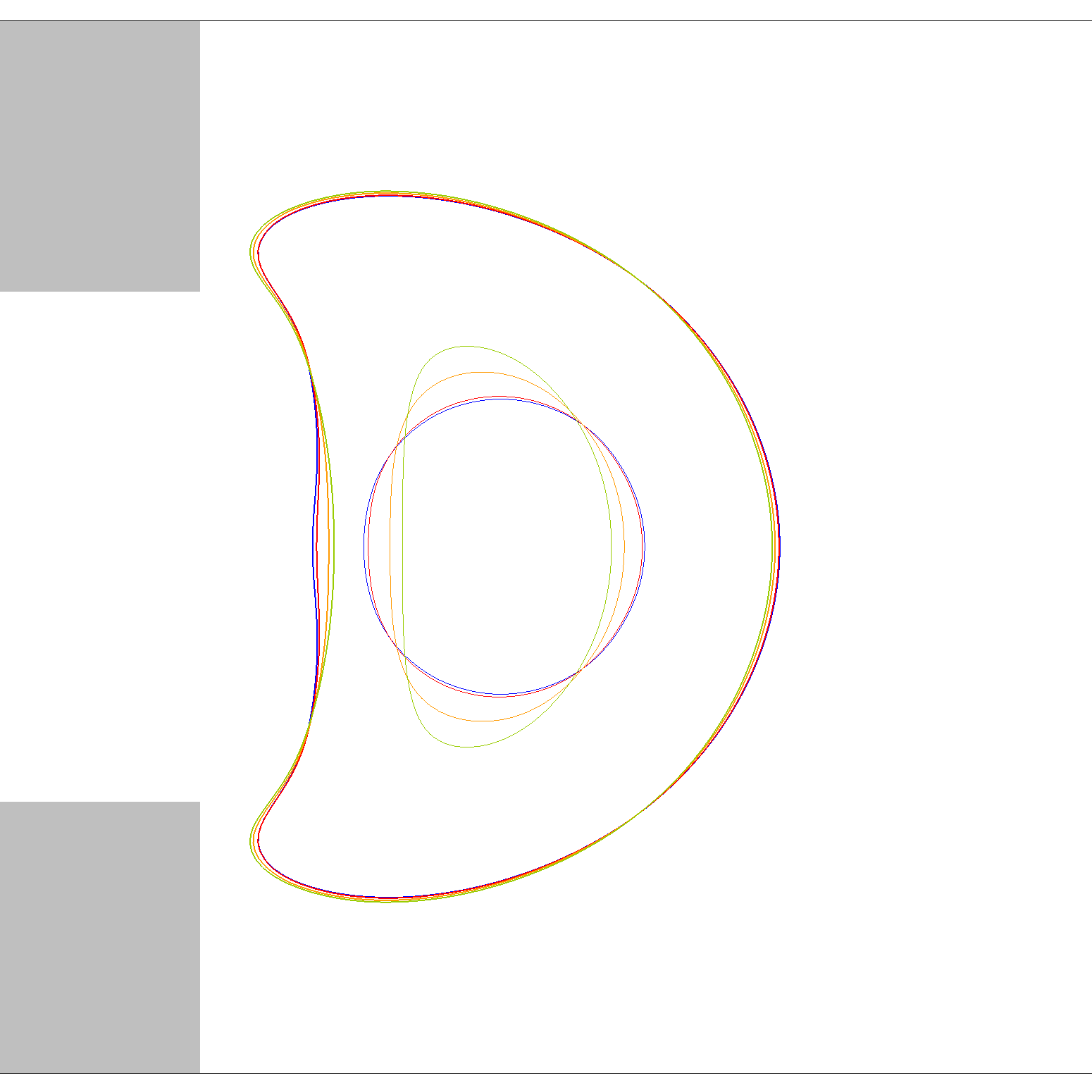}}
\caption{Capsule relaxation phenomenon: time evolution of the membrane and bulk profiles at different times for fixed radius ratio $a_b/a_m=0.5$ and variable elastic modulus of the bulk. The numerical results of TC16 are depicted in blue, TC17 in red, TC18 in orange, and TC19 in green. The walls of the square-section channel are represented in grey.}
\label{fig:relaxation_allradius05}
\end{figure}

Figure \ref{fig:square_relaxation_Lc1_fixedab} shows the time evolution of the $L_3/\ell$ characteristic length when varying the elastic modulus of the internal bulk, while keeping a constant radius $a_b/a_m=0.5$ (TC14, TC16, TC17, TC18, and TC19). 
Similar trends to Figure \ref{fig:square_relaxation_L1Lc1_fixedchi}b are followed when performing analysis on the bulk stiffness.
In particular, it should be noticed that the stiffer the internal bulk is the higher the minimum of $L_3/\ell$ gets.
As already observed in Figure \ref{fig:square_relaxation_L1Lc1_fixedchi}b, even though the capsule dynamics is highly affected by the
presence of the internal solid, the minimum value of $L_3/\ell$ always occurs at the same time.  \\
More details about the influence of an internal bulk on the capsule's shape is given in Figure \ref{fig:relaxation_allChi20}, 
where the capsule dynamics is depicted at different times for different configurations (TC15, TC16, TC20, and TC21).
For this case, an internal bulk with $\chi=2200$ is considered for the simulations. This brings the solid not to deform that much
during the relaxation process. However, the bulk's size remarkably influences the shape of the parachute that appears differently. 
In particular, the smaller the bulk is, the more the capsule resembles the phantom shape shown in 
Figure~\ref{fig:capsule_square_section_result_sk}a. 
Instead, when a bigger bulk is considered the shape of the parachute takes a more deformed profile.\\  
Finally, in Figure \ref{fig:relaxation_allradius05} a fixed ratio $a_b/a_m=0.5$ is considered, and 
the stiffness of the bulk varies from 110 to 2200.
In this case, the bulk now shows larger deformations because a lower elastic modulus is chosen. 
This brings the parachute to slightly modify its shape assuming a more, or less, elongated appearance. 
This also seems to be in line with the time evolution of $L_3/\ell$ presented in Figure \ref{fig:square_relaxation_Lc1_fixedab}. 

\section{Conclusion and discussion}\label{sec:conclusion}

In this paper, we have presented a numerical approach for the simulation of elastic capsules.
The numerical model is based on a fully Eulerian formulation for both the fluid and the deformable structures,
membrane and solid volume, that are here taken into account by the \revII{Navier--Stokes} equation by means of a source term
describing the elastic forces, which act on the fluid.
The fluid-structure system of equations considered herein describes both the forces related to the area
variation and shear by following the full membrane model introduced in \citep{Milcent2016}.
As mentioned above, the evaluation of the backward characteristics used to compute the elastic deformations can degrade
when complex simulations of capsules are addressed, especially when membrane are considered.
In this case, distortions of backward characteristics may occur within the membrane, causing a less reliable prediction
of the capsule dynamics.
The novelty of our method is based on coupling the Aslam extrapolation to improve the isolines of the backward
characteristics outside the capsule (as done in \citep{Deborde2020}) with the \textit{inner diffusion} approach to smooth
the internal isolines. Unfortunately, the good performances provided by the Aslam extrapolation in the outer zone does not
solve the same problem in the internal part.\\
The new algorithm is validated on several complex 3D configurations where the capsule dynamics is simulated and compared
to other numerical experiments present in the literature.
In particular, the challenges of these test cases are such that the proposed algorithm is essential to simulate
the correct evolution of the capsule.
First, we focus on the validation of the proposed model and method for classical test cases related to capsule dynamics.
We studied before the deformation of a capsule when immersed into a linear shear velocity field, which gives rise to the 
tank-treading motion. Then, we performed several simulations of a capsule in a square-section channel, which allows to study 
different parachute shapes that the capsule takes when highly compressed.
The numerical results obtained for different capillary numbers and constitutive laws have been compared with other numerical 
techniques present in the literature, obtaining a good agreement in all cases.
In~\ref{sec:circularshear}, we also present a grid refinement analysis for an academic test case to understand the 
needed level of refinement to achieve satisfying results. 
In particular, a modification of the circular shear benchmark introduced in~\citep{Milcent2016} is proposed by introducing 
a pressurization coefficient that allows, with the right constitutive law, the correct recovery of the spherical shape 
after its relaxation.
Finally, an original test case is presented where an elastic nucleus is considered within the thin membrane.
We provide interesting insights about the influence of a nucleus on the deformation of a capsule during its relaxation
after a compressed phase.

\revI{Future developments of our framework, in order to increase the range of applications that can be covered, will be dedicated to the simulations of multiple interacting capsules by introducing contact modeling techniques. A good starting point will be the work presented in~\cite{jedouaa2019efficient}. In particular, the Eulerian description of interfaces through level-sets gives direct access to normals and distances, which makes it very convenient to model contact between objects.}

\appendix

\section{Relaxation of a sheared elastic sphere}\label{sec:circularshear}

\begin{figure}
    \centering
    \includegraphics[width=0.45\textwidth]{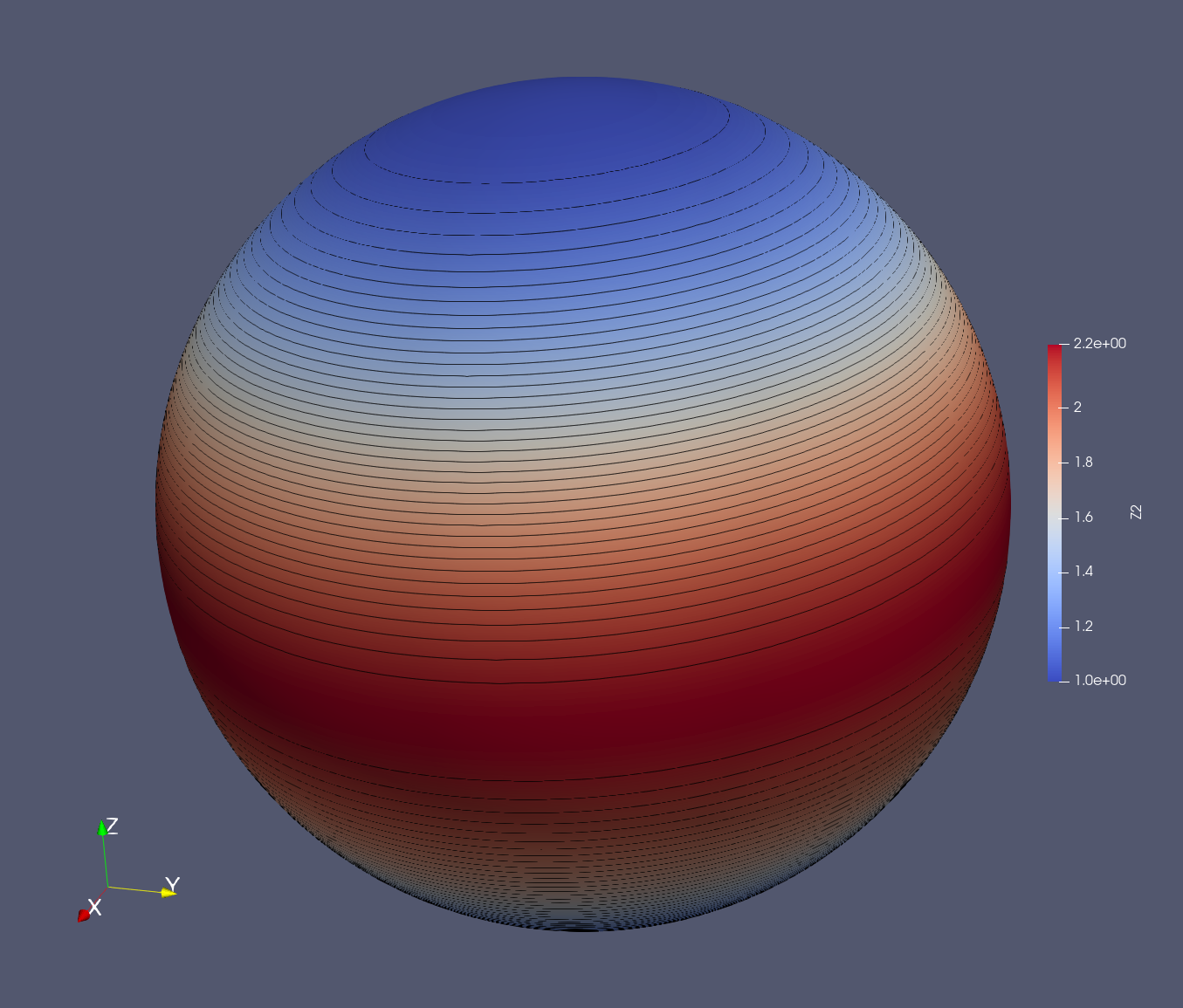}
    \caption{Relaxation of a sheared elastic sphere: initial condition of $Z_2$.}
    \label{fig:circular_shear_surface_t0}
\end{figure}

\begin{figure}
    \centering
    \subfloat[$t = 0.1 s$]{\includegraphics[width=0.3\textwidth]{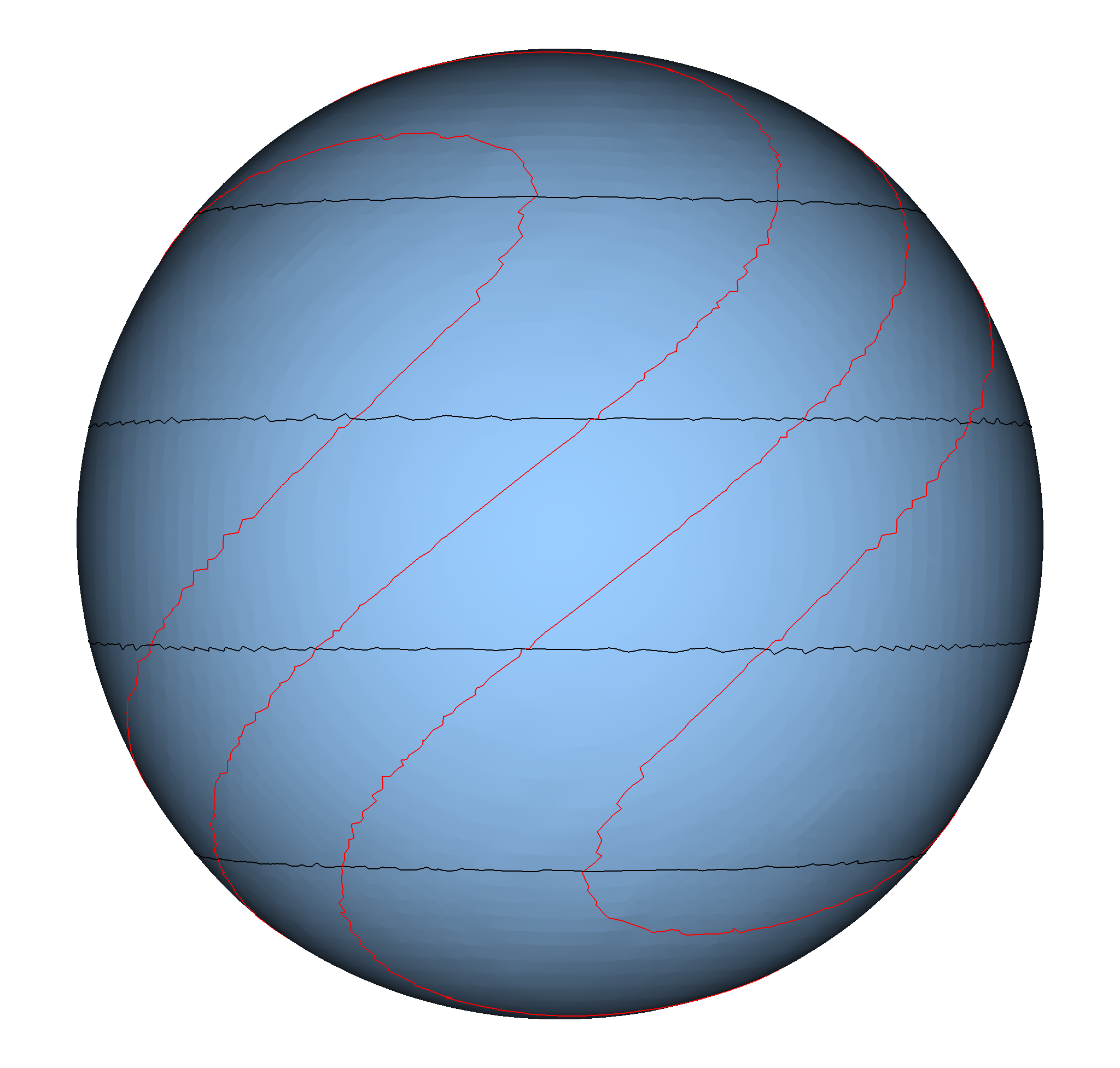}}
    \subfloat[$t = 0.5 s$]{\includegraphics[width=0.3\textwidth]{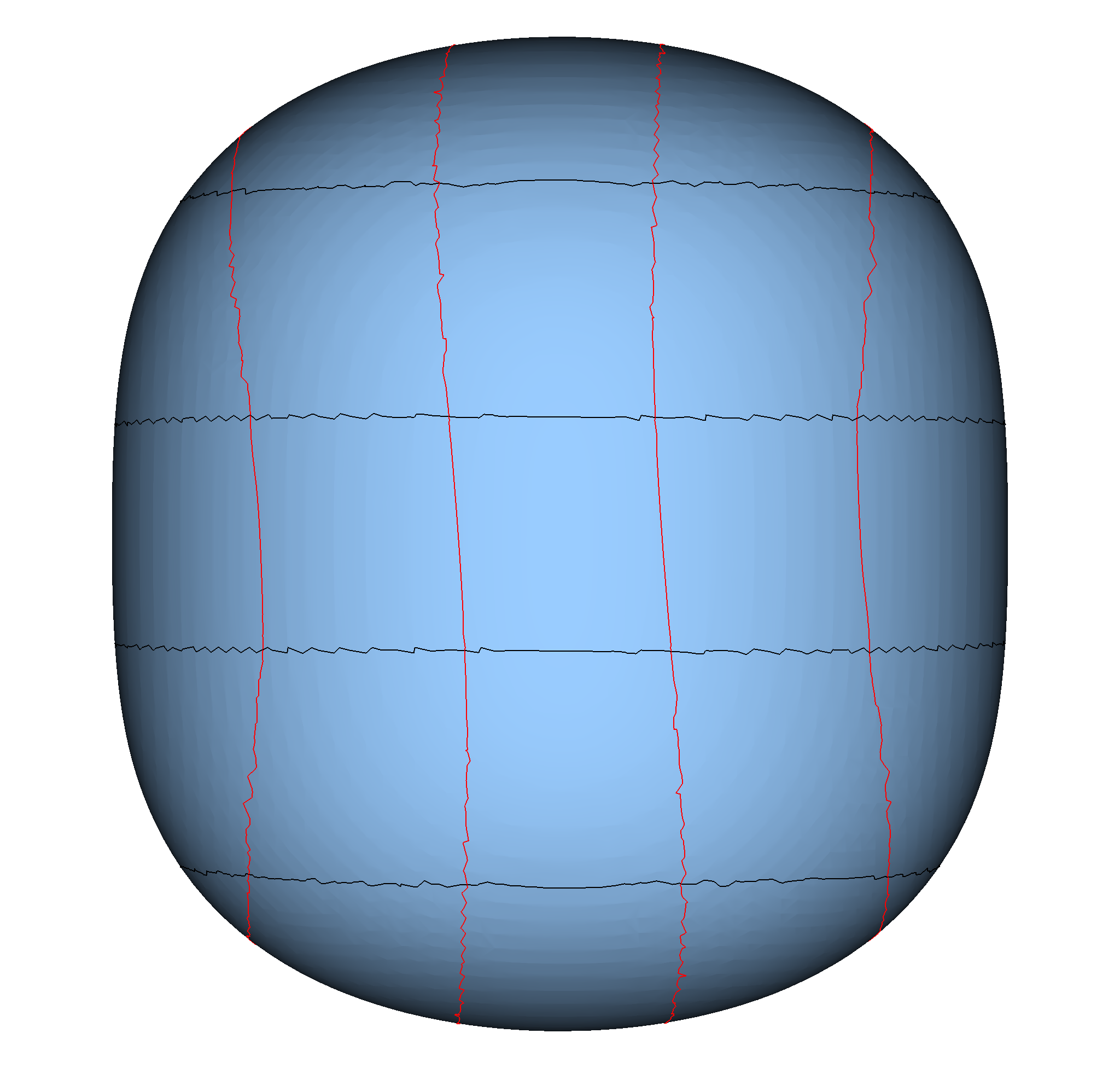}}
    \subfloat[$t = 1.2 s$]{\includegraphics[width=0.3\textwidth]{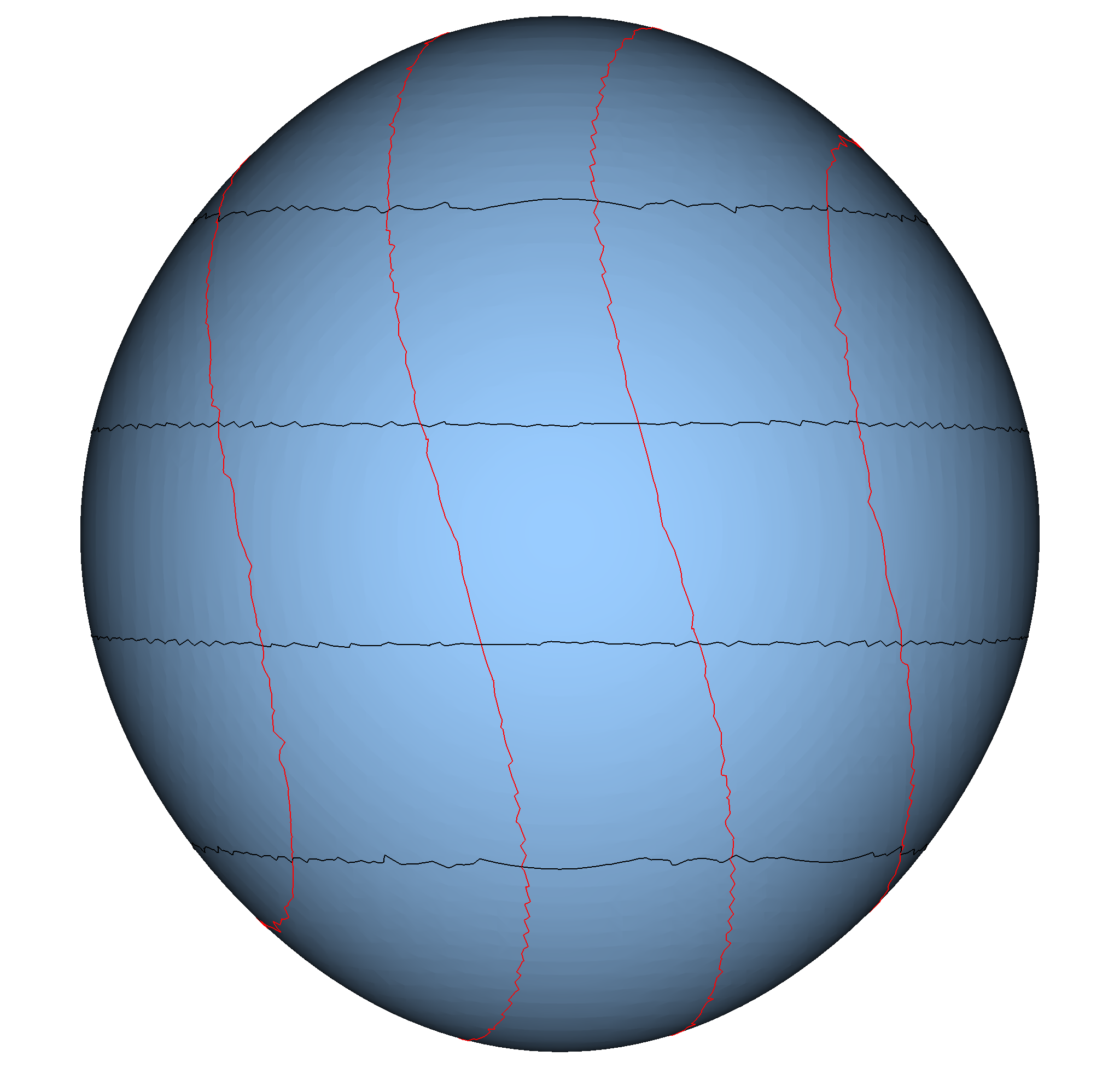}}\quad
    \subfloat[$t = 2 s$]{\includegraphics[width=0.3\textwidth]{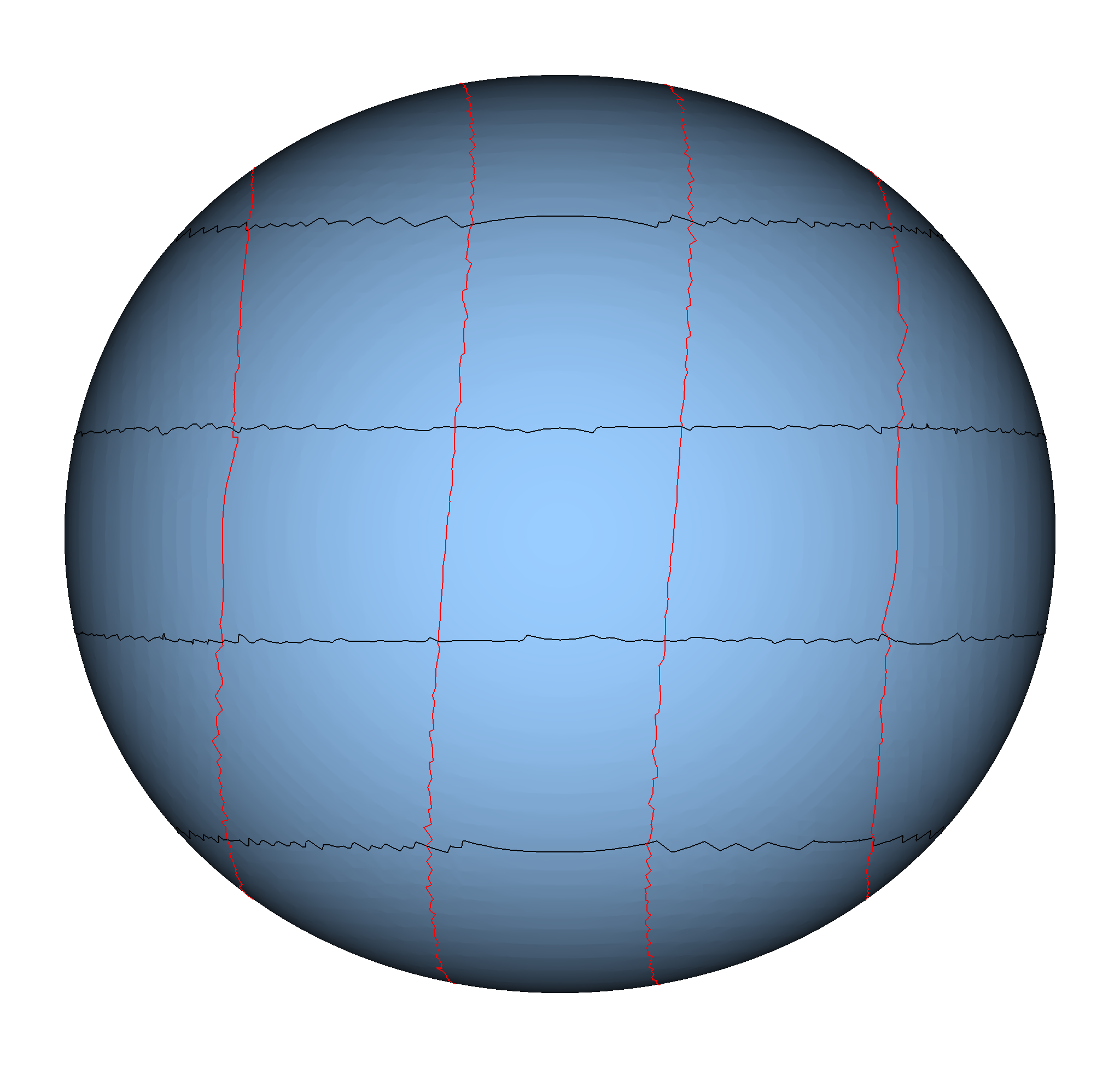}}
    \subfloat[$t = 3 s$]{\includegraphics[width=0.3\textwidth]{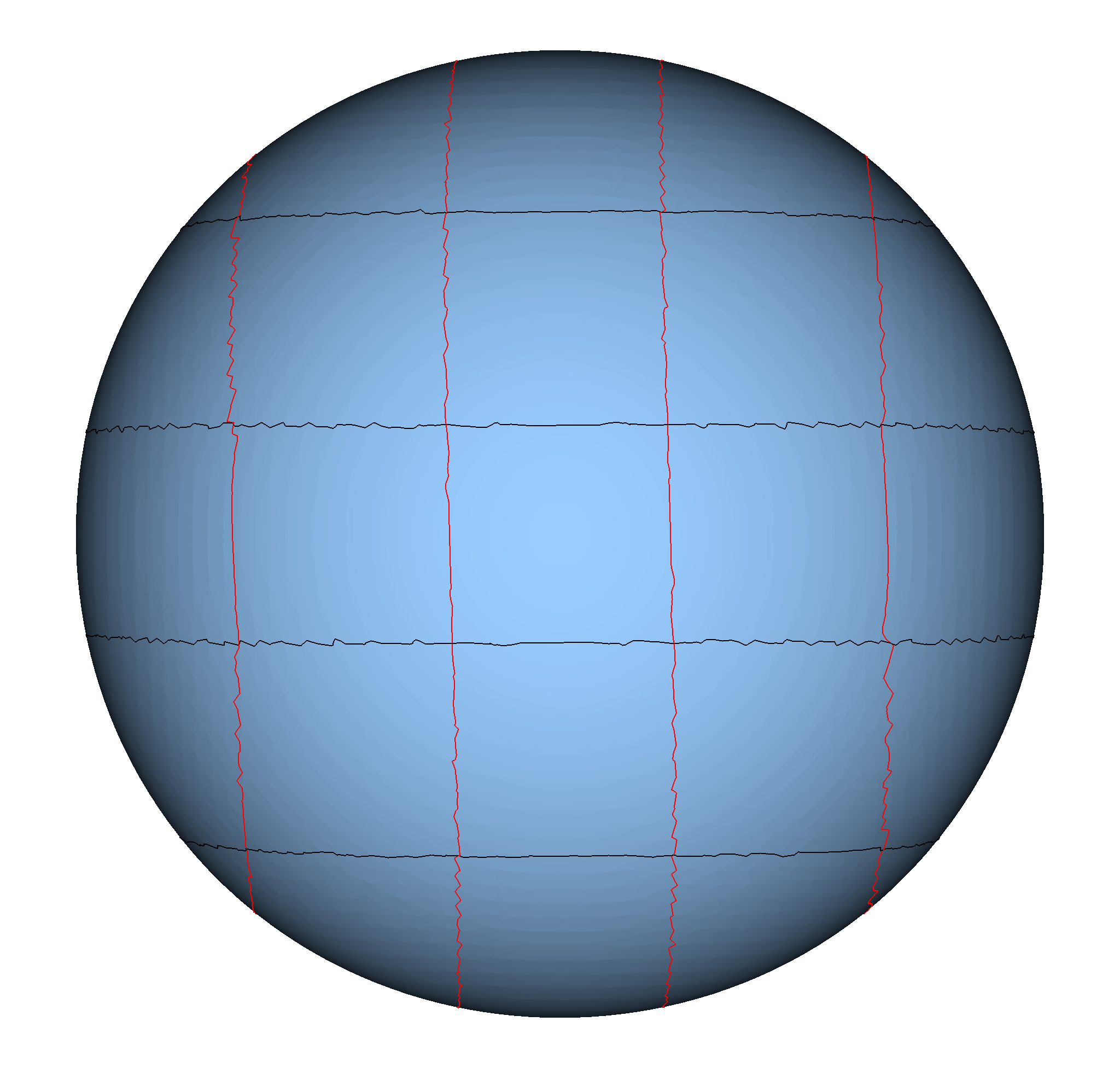}}
    \subfloat[$t = 5 s$]{\includegraphics[width=0.3\textwidth]{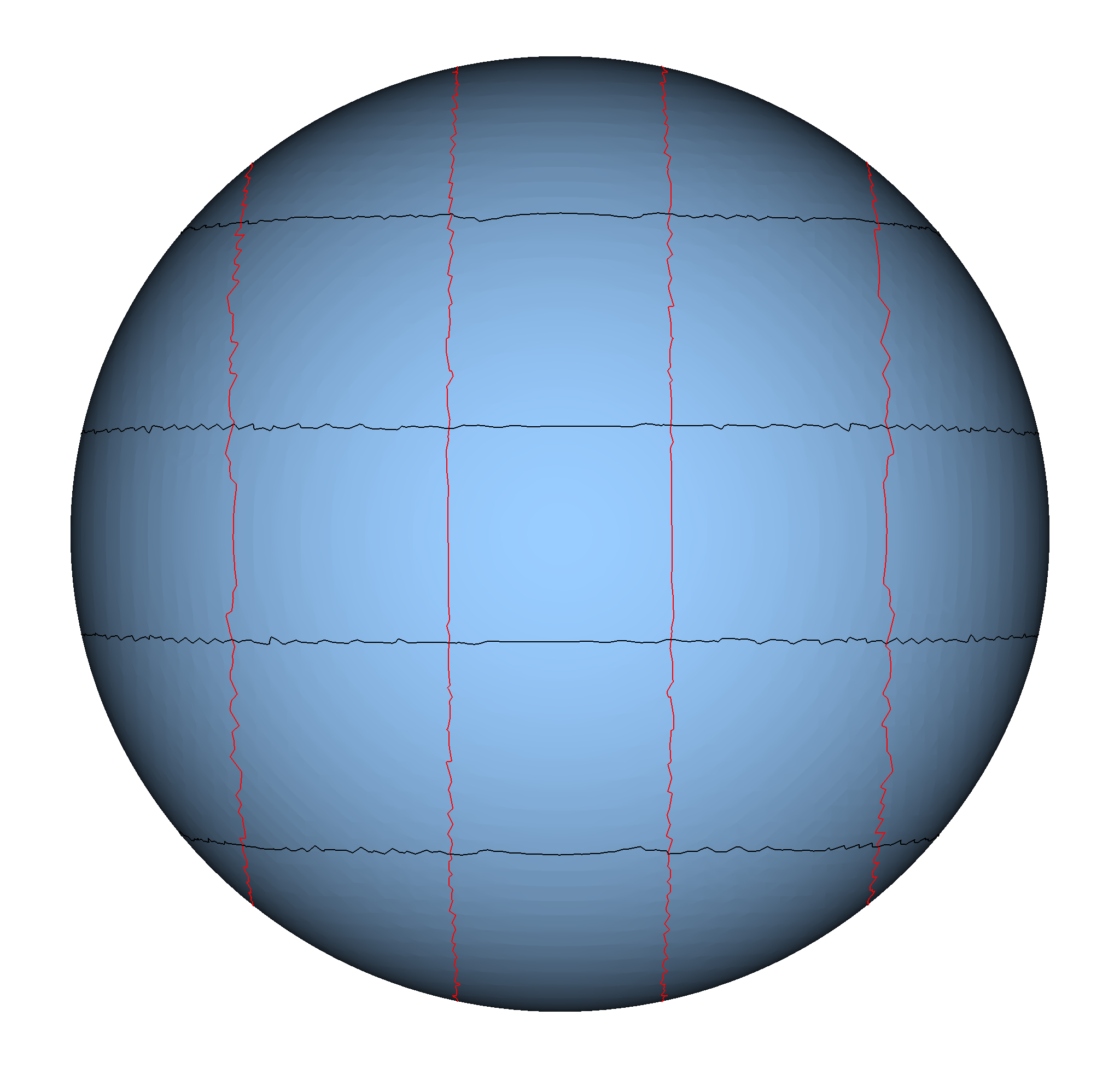}}
    \caption{Relaxation of a sheared elastic sphere: surface profile with the deformation isocontours $Y_1$ (red) and $Y_3$ (black) at different time steps.}
    \label{fig:circular_shear_surface_Y}
\end{figure}

\begin{figure}
\centering
\subfloat[$x$-radius]{
    \begin{tikzpicture}[
       scale=1.0,
        >=triangle 45,
        mydeco/.style = {decoration = {markings,
        mark = at position #1 with {\arrow{>}}}},
        baseline={(-2,0)}
        ]

        \begin{axis}[xlabel={Time ($s$)}, ylabel={$x$-radius /$a$},
        legend style={at={(0.8,0.3)},anchor=north}]
            \addplot[red,only marks,mark options={scale=0.5},mark=square] table[x index=0,y expr=\thisrowno{1}*2] {Circular_shear/radius_N32.txt};
            \addplot[green,only marks,mark options={scale=0.5},mark=triangle] table[x index=0,y expr=\thisrowno{1}*2] {Circular_shear/radius_N64.txt};
            \addplot[blue,smooth] table[x index=0,y expr=\thisrowno{1}*2] {Circular_shear/radius_N128.txt};
            \legend{$N=64$,$N=128$,$N=256$}
        \end{axis}

    \end{tikzpicture}
}
\subfloat[$z$-radius]{
    \begin{tikzpicture}[
       scale=1.0,
        >=triangle 45,
        mydeco/.style = {decoration = {markings,
        mark = at position #1 with {\arrow{>}}}},
        baseline={(-2,0)}
        ]

        \begin{axis}[xlabel={Time ($s$)}, ylabel={$z$-radius / $a$},
        legend style={at={(0.8,0.3)},anchor=north}]
            \addplot[red,only marks,mark options={scale=0.5},mark=square] table[x index=0,y expr=\thisrowno{2}*2] {Circular_shear/radius_N32.txt};
            \addplot[green,only marks,mark options={scale=0.5},mark=triangle] table[x index=0,y expr=\thisrowno{2}*2] {Circular_shear/radius_N64.txt};
            \addplot[blue,smooth] table[x index=0,y expr=\thisrowno{2}*2] {Circular_shear/radius_N128.txt};
            \legend{$N=64$,$N=128$,$N=256$}
        \end{axis}

    \end{tikzpicture}
}
    \caption{Relaxation of a sheared elastic sphere: evolution of the $x$-radius and $z$-radius for $N=\{64,128,256\}$.}
    \label{fig:circular_shear_xz_radius}
\end{figure}
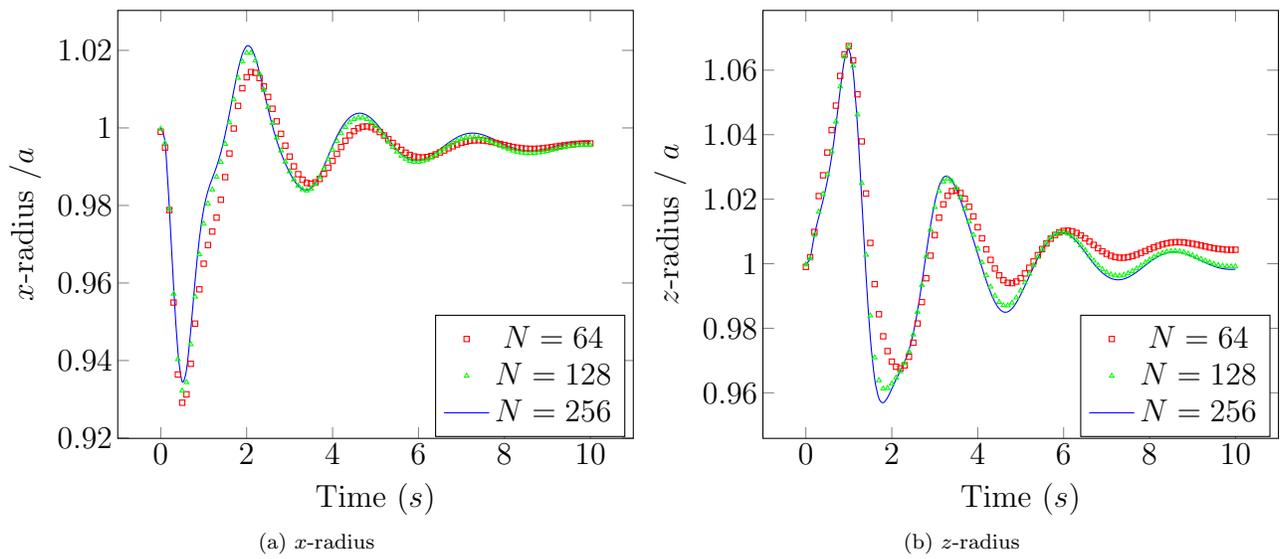

This academic test case is a modification of the sheared elastic sphere relaxation proposed in \citep{Milcent2016}.
Herein, we propose to use again this test case to validate the new method described in this paper by performing a mesh refinement analysis.
As it was shown in \citep{Milcent2016}, the relaxation of the pre-deformed sphere ends with
a final configuration that does no longer resemble an actual sphere. This is mainly related to the fact that the sphere was initialized
with no area variation ($Z_1=1$), which eventually brings the membrane to modify its aspect.\\
The simulation is setup by considering a thin membrane, modeled with the \revII{Evans--Skalak} constitutive law
\eqref{eq:law_evan_skalak} immersed within a viscous flow in the domain $[-1,1]^3$.
It should be noticed that also the constitutive law has changed with respect to the test case presented in \citep{Milcent2016}.
\revI{This change is related to the fact that the law used in \cite{Milcent2016}  does not represent any classical constitutive law, being the work mainly focused on the mathematical modeling of shear in the Eulerian framework.}
The physical parameters used for this test case are given in Table~\ref{tab:param_circular_shear}.
We impose zero velocity as initial condition and Neumann boundary conditions on all boundaries of the domain.
\begin{table}
    \centering
    \begin{tabular}{c|c|c|c|c}
          $\rho$ & $\mu$ & $G_s$ & $K_s$ & $a_m$ \\
         \hline
          1 & 0.01 & 0.1 & 1 & 0.5
    \end{tabular}
    \caption{Relaxation of a sheared elastic sphere: physical parameters.}
    \label{tab:param_circular_shear}
\end{table}
The initial conditions for the level-set field $\phi$ is given by a sphere of radius $a_m$,
\begin{equation}\label{eq:phisphere}
    \phi_0(x,y,z) = \sqrt{x^2 + y^2 + z^2} - a_m.
\end{equation}
The membrane is pre-deformed with a 3D circular shear until a fictitious time $t_0=\pi$.
This corresponds to the following initial conditions on $Y$,
\begin{equation}
    Y_0(x,y,z) = \frac{1}{1+\alpha}
        \begin{pmatrix}
         x\cos(t_0 z)+y\sin(t_0 z)\\
         y\cos(t_0 z)-x\sin(t_0 z)\\
         z
        \end{pmatrix}
\end{equation}
where $\alpha$ is the pressurization coefficient.
Following the Equation~\eqref{Z1_Z2b}, the initial area and shear variations reads,
\begin{equation}
Z_{1,0} = (1+\alpha)^2,\quad Z_{2,0} = 1+\frac{t_0^2(x^2+y^2)^2}{2(x^2+y^2+z^2)}.
\end{equation}
This introduces a shear variation that is equal to zero at the poles and varies along the $z$-axis (see Figure~\ref{fig:circular_shear_surface_t0}).
It should be noticed that, $\alpha$ does not influence the shear variation which stays the same,
but introduce an area variation on the initial conditions.
The coefficient $\alpha$ has been introduced to obtain a sphere shape after the relaxation process.
Moreover, when taking $\alpha=0$ we obtain the same configuration analyzed in~\citep{Milcent2016}.\\
The computational domain is discretized using a Cartesian mesh with $N=\{64,128,256\}$ cells in each direction.
The time step chosen for the simulations varies depending on the mesh refinement:
$\Delta t=4\times10^{-3}$ for $N=64$, $\Delta t=2\times10^{-3}$ for $N=128$ and $\Delta t=10^{-3}s$ for $N=256$.
In this simulation the pressurization coefficient $\alpha$ is set to 0.05.\\
The numerical results at different time steps are presented in Figure \ref{fig:circular_shear_surface_Y}, showing the deformation
isocontours until the new equilibrium is reached. At the beginning of the simulation, only the force related to the shear coefficient
is acting on the membrane. However, the membrane inertia introduces a shape modification that brings into play also the force term
related to the area variation.
As expected, the shear deformation gives rise to a notable deformation and inertia, along both the $x$ and $z$ axes, such that the
membrane begins to swing to one side and then to the other.
From $t=3s$, the membrane starts stabilizing and slightly oscillating around the steady state
where $Z_2$ is uniformly equal to one, and $Z_1=Z_{1,0}$. Because of the incompressibility condition, the internal fluid is going to
be characterized by the same volume, that translates into an area variation at the equilibrium equal to the initial one.
Having introduced the pressure coefficient $\alpha$, the equilibrium configuration has an area variation greater than one that implies
an elastic force at the interface. The pressure jump resulting from the elastic force is the reason why we are able to return to
an actual sphere configuration at the equilibrium (which was not achieved in~\citep{Milcent2016} where $\alpha$ was virtually set to zero).
In order to provide more insights about these oscillations, in Figure \ref{fig:circular_shear_xz_radius}, we plotted the evolution of the horizontal and vertical radii. These curves also give an idea about the mesh refinement needed to achieve good enough results.
The dimensionless pressure along the $x$ and the $z$ axes is also studied, in Figures \ref{fig:circular_shear_px} and
\ref{fig:circular_shear_pz} to provide more quantitative results on the mesh convergence.
It is possible to notice that different results, with respect to those in \citep{Milcent2016}, are found here because of aforementioned differences in the
initial pressurization of the membrane and elastic constitutive law.
\begin{figure}
\centering
\subfloat{
\begin{tikzpicture}
	\begin{axis}[
	   xlabel={$x/a$},
           ylabel={$pa/G_s$},
	   legend style={nodes={scale=0.8, transform shape},at={(0.5,0.3)},anchor=north},
	   width=.45\textwidth
	]
        \addplot[red,only marks,mark options={scale=0.5},mark=square] table[x expr=\thisrowno{0}*2,y expr=\thisrowno{1}*5] {Circular_shear/n64/Circular_shear_n32_pressure_x_t01.txt};
        \addplot[green,only marks,mark options={scale=0.5},mark=triangle] table[x expr=\thisrowno{0}*2,y expr=\thisrowno{1}*5] {Circular_shear/n128/Circular_shear_n64_pressure_x_t01.txt};
        \addplot[blue,smooth] table[x expr=\thisrowno{0}*2,y expr=\thisrowno{1}*5] {Circular_shear/n256/Circular_shear_n128_pressure_x_t01.txt};
	\legend{$N=64$,$N=128$,$N=256$}
	\end{axis}
\end{tikzpicture}}\quad
\subfloat{
\begin{tikzpicture}
        \begin{axis}[
           xlabel={$x/a$},
           ylabel={$pa/G_s$},
           legend style={nodes={scale=0.8, transform shape},at={(0.5,0.3)},anchor=north},
           width=.45\textwidth
        ]
	\addplot[red,,only marks,mark options={scale=0.5},mark=square]   table[x expr=\thisrowno{0}*2,y expr=\thisrowno{1}*5] {Circular_shear/n64/Circular_shear_n32_pressure_x_t12.txt};
	\addplot[green,only marks,mark options={scale=0.5},mark=triangle]  table[x expr=\thisrowno{0}*2,y expr=\thisrowno{1}*5] {Circular_shear/n128/Circular_shear_n64_pressure_x_t12.txt};
	\addplot[blue,smooth] table[x expr=\thisrowno{0}*2,y expr=\thisrowno{1}*5] {Circular_shear/n256/Circular_shear_n128_pressure_x_t12.txt};
        \legend{$N=64$,$N=128$,$N=256$}
        \end{axis}
\end{tikzpicture}}
\caption{Relaxation of a sheared elastic sphere: dimensionless pressure $pa/G_s$ along the $x$-axis at time $t=0.1 s$ (left), and $t=1.2$ (right) for $N=\{64,128,256\}$.}
\label{fig:circular_shear_px}
\end{figure}
\begin{figure}
\centering
\subfloat{
\begin{tikzpicture}
        \begin{axis}[
           xlabel={$z/a$},
           ylabel={$pa/G_s$},
           legend style={nodes={scale=0.8, transform shape},at={(0.5,0.3)},anchor=north},
           width=.45\textwidth
        ]
        \addplot[red,only marks,mark options={scale=0.5},mark=square] table[x expr=\thisrowno{0}*2,y expr=\thisrowno{1}*5] {Circular_shear/n64/Circular_shear_n32_pressure_z_t01.txt};
        \addplot[green,only marks,mark options={scale=0.5},mark=triangle] table[x expr=\thisrowno{0}*2,y expr=\thisrowno{1}*5] {Circular_shear/n128/Circular_shear_n64_pressure_z_t01.txt};
        \addplot[blue,smooth] table[x expr=\thisrowno{0}*2,y expr=\thisrowno{1}*5] {Circular_shear/n256/Circular_shear_n128_pressure_z_t01.txt};
        \legend{$N=64$,$N=128$,$N=256$}
        \end{axis}
\end{tikzpicture}}\quad
\subfloat{
\begin{tikzpicture}
        \begin{axis}[
           xlabel={$z/a$},
           ylabel={$pa/G_s$},
           legend style={nodes={scale=0.8, transform shape},at={(0.5,0.3)},anchor=north},
           width=.45\textwidth
        ]
        \addplot[red,,only marks,mark options={scale=0.5},mark=square]   table[x expr=\thisrowno{0}*2,y expr=\thisrowno{1}*5] {Circular_shear/n64/Circular_shear_n32_pressure_z_t12.txt};
        \addplot[green,only marks,mark options={scale=0.5},mark=triangle]  table[x expr=\thisrowno{0}*2,y expr=\thisrowno{1}*5] {Circular_shear/n128/Circular_shear_n64_pressure_z_t12.txt};
        \addplot[blue,smooth] table[x expr=\thisrowno{0}*2,y expr=\thisrowno{1}*5] {Circular_shear/n256/Circular_shear_n128_pressure_z_t12.txt};
        \legend{$N=64$,$N=128$,$N=256$}
        \end{axis}
\end{tikzpicture}}
\caption{Relaxation of a sheared elastic sphere: dimensionless pressure $pa/G_s$ along the $z$-axis at time $t=0.1 s$ (left), and $t=1.2$ (right) for $N=\{64,128,256\}$.}
\label{fig:circular_shear_pz}
\end{figure}


\bibliographystyle{cas-model2-names}

\bibliography{bib.bib}

\end{document}